\documentclass[bibyear]{aa}

\usepackage{graphicx}
\usepackage{txfonts}
\usepackage{natbib,twoopt}
\usepackage{hyperref} 
\bibpunct{(}{)}{;}{a}{}{,}             

\usepackage{amsmath,bm}	
\usepackage{orcidlink}

\begin{document}

\title{Thermal emission from bow shocks. III. Variable diffuse X-ray emission from stellar-wind bow shocks driven by dynamical instabilities}
\titlerunning{X-ray emission from wind-driven bow shocks}

\author{
  Jonathan~Mackey~\orcidlink{0000-0002-5449-6131}\inst{1}\thanks{\email{jmackey@cp.dias.ie}}
  \and 
  Arun~Mathew~\orcidlink{0000-0001-9896-4243}\inst{1}
  \and 
  Ahmad~A.~Ali~\orcidlink{0000-0001-5189-4022}\inst{2}
  \and 
  Thomas~J.~Haworth~\orcidlink{0000-0002-9593-7618}\inst{3} \and 
  Robert~Brose~\orcidlink{0000-0002-8312-6930}\inst{4,5,1}
  \and
  Sam~Green~\orcidlink{0000-0003-1129-4676}\inst{6}
  \and  
  Maria~Moutzouri~\orcidlink{0000-0002-6501-090X}
  \and 
  Stefanie~Walch~\orcidlink{0000-0001-6941-7638}\inst{2}
}

\institute{
  Astronomy \& Astrophysics Section, School of Cosmic Physics, Dublin Institute for Advanced Studies, DIAS Dunsink Observatory, Dublin D15 XR2R, Ireland
\and
  I. Physikalisches Institut, Universit\"{a}t zu K\"{o}ln, Z\"{u}lpicher Str. 77, 50937 K\"{o}ln, Germany
\and
  Astronomy Unit, School of Physics and Astronomy, Queen Mary University of London, London E1 4NS, UK
\and
  Institute of Physics and Astronomy, University of Potsdam, 14476 Potsdam-Golm, Germany
\and
  School of Physical Sciences and Centre for Astrophysics \& Relativity, Dublin City University, Glasnevin, D09 W6Y4, Ireland
\and
  ARC Centre of Excellence for the Weather of the 21st Century, University of New South Wales, Sydney, Australia
}

\date{Received 2025-01-10; Revised 2025-03-05; Accepted 2025-03-09}

\abstract
{X-ray emission from wind-driven bow shocks is both difficult to measure and predict, but may give important insights into the energy budget of the hot phase of the interstellar medium (ISM) by quantifying mixing at the interface between hot and warm gas phases.}
{We investigate the effect of magnetic fields and numerical resolution on predicted X-ray emission and other observable properties of bow shocks, to study convergence properties and assess robustness of predicted observables from simulations.}
{A suite of 2D and 3D hydrodynamic and magnetohydrodynamic simulations of bow shocks were run and analysed to generate synthetic emission maps and light curves in X-ray and infrared emission.}
{Resolving the Kelvin-Helmholtz (KH) instability at the wind-ISM contact discontinuity is crucial for obtaining converged results and for predicting X-ray emission and the properties of the hot shocked wind.
When sufficient spatial resolution is used, we measure time variation of X-ray emission of at least an order of magnitude on a timescale comparable to the advection timescale of the wake downstream from the bow shock.
Good correspondence is found between 2D and 3D simulations with comparable resolution, and 3D simulations can achieve the required resolution with reasonable computing resources.
Development of the KH instability is inhibited for shear flows parallel to the ISM magnetic field, compared with what is seen in the perpendicular direction, resulting in synthetic IR emission maps of bow shocks that are smooth when seen from one perspective but show strong distortions from another.}
{
Measuring the X-ray morphology and luminosity in bow shocks may be useful for constraining mixing and energy-transfer rates between hot and warm gas phases of the ISM.
Dynamical instabilities at the wind-ISM interface are a crucial ingredient in determining the properties of the hot-gas phase in stellar bow-shocks, in particular to capture its time dependence.
}


\maketitle


\section{Introduction}
\label{sec:introduction}
Stellar-wind bubbles form around massive stars because the ram pressure of the radially expanding wind displaces the ISM outwards, creating a hot, shocked-wind bubble surrounded by swept-up interstellar gas \citep{DysDeV72, WeaMcCCas77}.
A key issue in the dynamical evolution of wind-driven nebulae is whether the bubble expansion is energy conserving or momentum conserving.
\citet{WeaMcCCas77} assumed an energy-conserving expansion (i.e.\ no energy is dissipated or escapes from the bubble to the ISM) and found that the bubble radius scales with time as $t^{3/5}$.
However, if energy rapidly leaks out of the bubble, the expansion is driven only by the radial momentum of the wind, and the radius scales as $t^{1/2}$.
An internal inconsistency of the Weaver model is that it assumes thermal conduction at the wind-ISM interface (the contact discontinuity, hereafter CD) efficiently facilitates heat transfer across this interface and out of the hot bubble, while simultaneously assuming the energy-conserving solution.
Conduction produces a layer of intermediate density and temperature gas that emits strongly in X-rays with energy $kT\sim0.1-1$\,keV and in UV lines of ions such as O$^{5+}$.
The interior of the bubble cannot cool effectively because it has very low density and high temperature, and so energy loss is almost entirely through the outer boundary of the hot bubble at the CD, extensively studied by \citet{LanOstKim21a, LanOstKim24}.
It is therefore crucial to accurately model the CD and wind-ISM mixing processes in order to correctly predict the X-ray emission and energy content of stellar-wind bubbles.

Simulations of wind bubbles around static stars, both including and excluding thermal conduction, have shown that much of the bubble energy is lost through the mixing of wind and ISM material across the CD.
This results in an intermediate-temperature gas $T\sim10^5-10^6$\,K that cools very strongly in UV lines \citep[e.g.][]{FreHenYor06, ToaArt11}.
Mixing may arise from numerical diffusion, shown to lead to strong over-cooling in superbubble simulations \citep{MacMcCNor89} if the diffusion is too large.
Possible physical sources of mixing are the evaporation of dense clumps \citep[i.e. mass-loading][]{McKVanLaz84, Art12}, thermal conduction \citep{CowMcK77, WeaMcCCas77}, and turbulent mixing \citep{BegFab90, SlaShuBeg93, EsqBenLaz06}.

Turbulent mixing, essentially entrainment of cold gas into the hot phase by dynamical instabilities such as Rayleigh-Taylor (RT) or Kelvin-Helmholtz (KH), was investigated by \citet{BegFab90} at hot-cold gas interfaces in clusters of galaxies.
They showed that efficient mixing can create and maintain a layer of intermediate temperature gas that emits in the far UV and can explain, for example, [\ion{O}{VI}] lines observed around cold clouds embedded in the intracluster medium.
\citet{SlaShuBeg93} modelled the properties of turbulent mixing layers between hot and cool gas, initiated specifically by shear flows at the interface, 
predicting that up to 20\% of the supernova input power to the ISM is processed through mixing layers.

\citet{EsqBenLaz06} studied shear-driven turbulent mixing with 3D magnetohydrodynamics (MHD) simulations of a perturbed, planar, shear-layer with periodic boundary conditions.
They found that the development of the mixing layer is slow, with a timescale of $\approx 1$\,Myr, although this must be related to the large box-size and low resolution of the simulations.
A similar setup was revisited by \citet{JiOhMas19}, who ran a suite of 3D MHD simulations with varying gas metallicity (and hence cooling rate), shear velocity and gas density to test the predictions of \citet{BegFab90} for mixing-layer thickness and mean temperature against simulation results, finding a clear disagreement.
This was explored further by \citet{FieDruOst20}, who developed a fractal model for mixing at shear interfaces, tested against 3D hydrodynamic simulations of planar shear layers.
They noted that mixing is expected to be enhanced in environments with short cooling time, large shear velocity, and large density ratio across the CD, of which at least two criteria are satisfied at the CD of bow shocks around hot stars.

\citet{TanOhGro21} compared the mixing-layer models to the problem of turbulent combustion, finding strong parallels that could explain the scaling of energy dissipation on the mixing layer as a function of the other parameters of the system (e.g.\ density contrast, relative strength of radiative cooling and thermal conduction).
\citet{TanOhGro21} also explored why numerical convergence may be achieved, at least in terms of energy dissipation rate in the mixing layer, even when all of the relevant length scales are not resolved in simulations.
Closely related, the fractal model of \citet{LanOstKim21a, LanOstKim21b, LanOstKim24} for turbulent mixing and energy loss at the wind-ISM interface of wind bubbles has many similarities to the shear-driven mixing models.
It may be less applicable to the bow shocks that are the focus of this work because in bow shocks the CD has a strong shear flow along most of its surface area except for the stagnation point at the apex of the bow shock.

Hydrodynamic simulations of bow shocks that include thermal conduction show a relatively broad and laminar mixing layer \citep{ComKap98, MeyMacLan14}, although \citet{MeyMigKui17} showed that the inclusion of an ISM magnetic field strongly inhibits conduction across the CD because the field lines are draped along the CD on account of the stellar motion through the ISM.
\citet{MeyMigKui17} found that MHD calculations gave significantly weaker X-ray emission from bow shocks than hydrodynamics (HD) calculations (their fig.~8), presumably because the mixing layer is much less extended when heat conduction is suppressed by the magnetic field.

\citet{MacGvaMoh15} investigated asymmetric stellar wind bubbles for stars moving with velocity, $v_\star \in [2,16]$\,km\,s$^{-1}$, that is, both subsonic and supersonic motion through the surrounding H~\textsc{ii} region.
They found in all cases that the KH instability raised waves on the wind-ISM interface that induced strong mixing in the downstream region.
Most of the input energy to the bubble was radiated by gas with $T<10^5$\,K, which would result in relatively weak X-ray emission but strong UV line emission; predictions were made for the emission from the Galactic bubble RCW\,120.
Subsequently, \citet{TowBroGar18} reported a detection of diffuse X-ray emission from RCW\,120; the X-ray data were re-analysed by \citet{LuiAndSch21} who found a luminosity of $L_\mathrm{X}\approx2\times10^{32}$\,erg\,s$^{-1}$, similar to the prediction of $10^{32}$\,erg\,s$^{-1}$ from gas with $T>10^6$\,K by \citet{MacGvaMoh15}.

\citet{GreMacHaw19} modelled the Bubble Nebula, NGC\,7635, as a bow shock produced by wind-ISM interaction from the runaway star BD+60$^\circ$\,2522, using HD simulations to predict the soft and hard X-ray emission both as a total luminosity and in emission maps for comparison with observations.
These predictions were tested by \citet{ToaGueTod20} with \textit{XMM-Newton} observations, and shown to be a significant over-prediction based on the estimated upper limits on diffuse X-ray emission.
Subsequently, \citet{GreMacKav22} made 3D MHD simulations of the bow shock of $\zeta$ Ophiuchi, comparing with X-ray observations from the \textit{Chandra X-ray Observatory} \citep{ToaOskGon16}.
In contrast to the results for NGC\,7635, here the simulations under-predicted the X-ray emission compared to the observational detection by a factor of a few.
In both cases there are uncertainties with the observations: for the case of NGC\,7635 related to the distance and foreground absorbing column of gas and dust, and for $\zeta$ Ophiuchi in the separation of stellar and diffuse X-ray photons due to pile-up.
Nevertheless, it is puzzling that both simulations give results inconsistent with observations but in the opposite sense in each case.
\citet{GreMacKav22} also showed that the predicted X-ray luminosity was also somewhat dependent on numerical resolution.
Other 3D MHD simulations of bow shocks around hot stars \citep{BaaSchKle21, BaaSchKle22} did not investigate X-ray emission, being focussed mainly on the shocked ISM and its emissivity.

This motivates us to undertake a detailed study of simulated X-ray emission from bow shocks where we vary the numerical resolution, the equations solved (Euler vs MHD), the Riemann solver used, and the dimensionality (2D vs.\ 3D).
We aim to determine to what degree X-ray emission from bow shocks can be predicted by simulations and what level of agreement we should expect between simulations and observations.
A priori, the reasoning for studying these parameters is as follows:
\begin{enumerate}
\item Numerical resolution determines the diffusivity of the simulation, thereby affecting the mass and volume of gas at intermediate temperatures at the wind-ISM interface;
\item Similarly the choice of the Riemann solver also changes the diffusivity of the integration scheme;
\item Magnetic fields are known to suppress the KH instability \citep{FraJonRyu96}, possibly affecting the degree of mixing produced by inhibiting development of turbulence;
\item Instabilities behave differently in 2D compared with 3D because the symmetry imposed by 2D simulations restricts gas motion to only certain types of flow; this may affect numerical mixing and, in particular, its time dependence; and
\item \citet{TanOhGro21} showed that numerically converged results can be obtained for shear-driven mixing layers even when the spatial resolution is not sufficient to resolve all of the relevant length-scales.
\end{enumerate}

In section~\ref{sec:methods} we introduce the numerical methods used, the suite of simulations that were run, and the post-processing methods used to analyse the datasets and generate synthetic emission maps.
Our results are presented in section~\ref{sec:results}, starting with a comparison of simulations with different Riemann solvers and resolutions, comparing HD and MHD simulations, and making a detailed analysis of 3D MHD simulations, and the time dependence of their X-ray and infrared emission.
Results are discussed in the context of previous work in section~\ref{sec:discussion} and our conclusions are reported in section~\ref{sec:conclusions}.

\section{Methods and initial conditions}
\label{sec:methods}

\begin{figure}
	\includegraphics[width=\columnwidth]{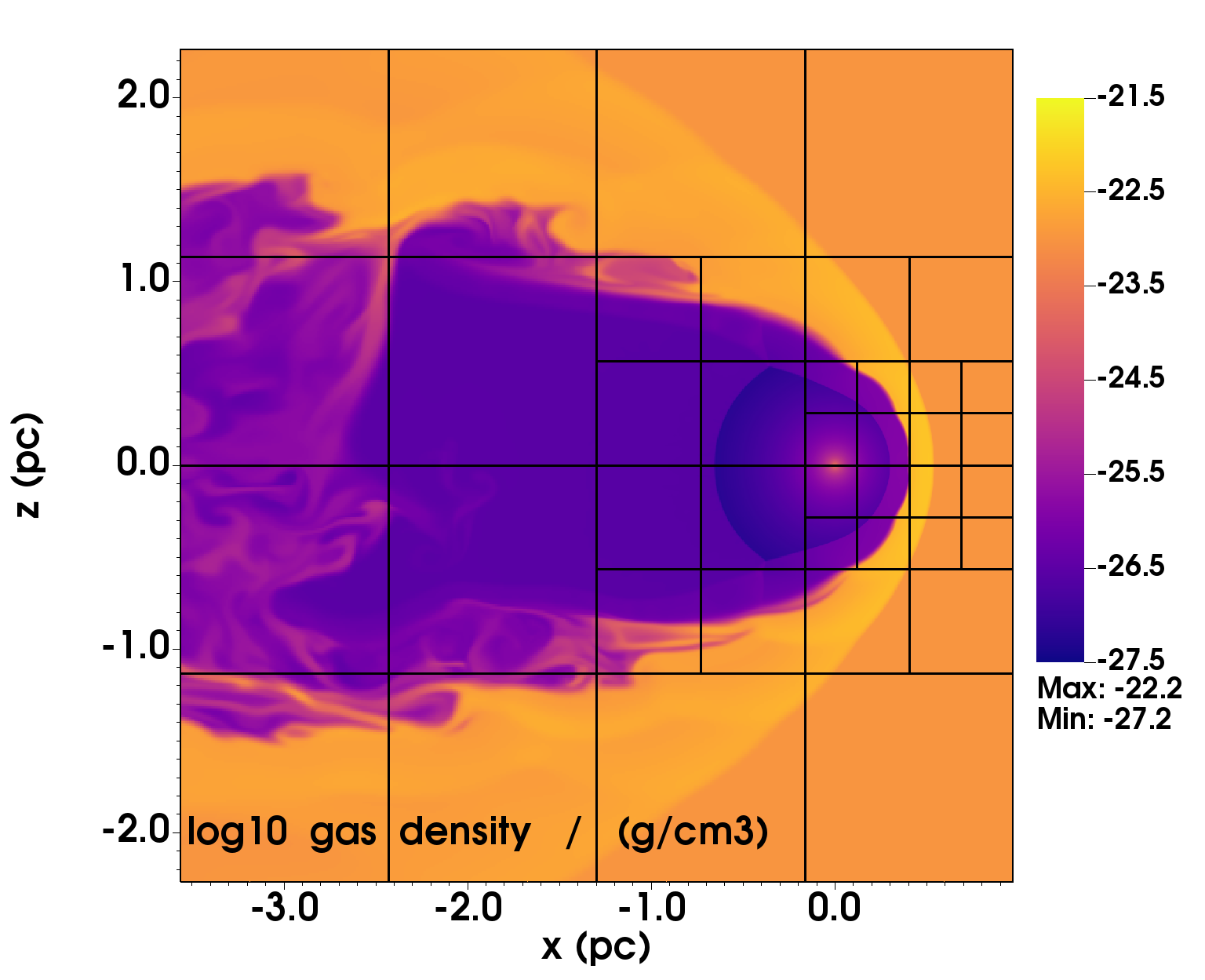}
    \caption{Slice of gas density through 3D simulation of a bow shock, in the $x$-$z$ plane.  The grid refinement is shown with the black lines, where each block has an equal number of grid cells and blocks are divided among MPI processes.  Resolution is focussed on the apex of the bow shock.}
    \label{fig:grid}
\end{figure}

We used the finite-volume MHD code \textsc{pion} for the simulations presented here, with static mesh-refinement as described in \citet{MacGreMou21}.
Bow shocks around massive stars were simulated in 2D cylindrical coordinates in the $R$-$z$ plane with assumed rotational symmetry \citep{MacMohNei12, GreMacHaw19}, and in 3D Cartesian coordinates \citep{MacGreMou21, GreMacKav22}, using a reference frame where the star is at rest and the ISM flows past the star in the -$\hat{\bm{x}}$ direction with the relative velocity between the runaway star and the ISM, $v_\star$.
Static mesh-refinement was used with three levels of refinement and a factor of 2 resolution increase with each level, focussed on the +$\hat{\bm{x}}$ domain boundary, chosen such that both the star and the apex of the bow shock are contained on the finest grid level.
An example grid from a 3D simulation is shown in Fig.~\ref{fig:grid}.

In 3D, the simulations were set up with a cubic domain with range $L_\mathrm{box} = 1.4\times10^{19}$ cm ($\approx 4.5$\,pc) in each dimension.
The $x$-domain is $x\in[-1.1,0.3]\times10^{19}$\,cm and the other two dimensions have $\{y,z\}\in[-0.7,0.7]\times10^{19}$\,cm.
Centred on $[0.3,0,0]\times10^{19}$\,cm, the level 1 grid has a domain size half of the level 0 grid in each dimension, and the level 2 grid is again half the size.
In 2D only the $R>0$ half-plane was simulated, with an axisymmetric boundary condition at the $R=0$ coordinate singularity, but otherwise the domain is the same as in 3D except that the relative motion between the star and the ISM is along the $\hat{\bm{z}}$ axis.

Optically thin radiative heating and cooling was assumed, following \citet{GreMacHaw19, GreMacKav22}, which is appropriate for the diffuse ISM conditions simulated here.
The \textsc{pion} cooling model 8 was used, which treats the ISM as photoionised by an extreme-UV (EUV, photon energy $>13.6$\,eV) radiation source with heating rate per photoionisation of 5\,eV, typically for the radiation field of an O star.
Assuming photoionisation equilibrium, the hydrogen photoionisation rate is equal to the Case B recombination rate, and the heating rate was obtained from this.
For the ISM and stellar properties discussed below, the star should have an ionising photon luminosity of $Q_0\approx3\times10^{48}\,\mathrm{s}^{-1}$, photoionising out to the Str\"omgren radius of around 15\,pc, much larger than the simulation domain and justifying our assumption of a fully ionised domain.
Radiative cooling is the sum of bremsstrahlung from H and He, collisional ionisation equilibrium metal-line cooling \citep{WieSchSmi09} and forbidden-line cooling from photoionised CNO ions around $10^4$\,K \citep{HenArtDeC09}.
The resulting equilibrium gas temperature is $\approx8300$\,K, independent of density.
Further details of the cooling prescription can be found in \citet{GreMacHaw19} and \citet{MacGreMou21}.

We are not aiming to model any particular bow shock, but rather to conduct a generic study that could be applicable to many systems.
As such we took typical wind properties for a late O-type star of about 20-25\,M$_\odot$, specifically a mass-loss rate of $\dot{M}=10^{-7}$\,M$_\odot$\,yr$^{-1}$ and a wind terminal velocity of $v_\infty=2000$\,km\,s$^{-1}$ \citep[e.g.][]{VinDeKLam00, PulVinNaj08}.
We also considered a typical runaway-star velocity through the ISM of $v_\star=30$\,km\,s$^{-1}$, similar to the space velocity measured for $\zeta$ Oph \citep{GvaLanMac12} and the driving star of the Bubble Nebula, BD+60$^\circ$2522 \citep{GreMacHaw19}.
Bow shocks are more likely to be detected for stars moving through a denser ISM because the surface brightness of the shocked region is larger \citep[see discussion in][]{KobSchBal17}, and so we considered a background ISM density of $\rho_0=10^{-23}$\,g\,cm$^{-3}$, or hydrogen number density of $n_\mathrm{H}\approx4.3\,$cm$^{-3}$.

\subsection{Boundary conditions}
\label{sec:methods:bcs}
The stellar wind was injected into the simulation domain from a sphere of radius 10-20 grid cells (depending on the simulation resolution).
The wind was injected at the terminal velocity because the wind boundary (radius $\approx10^{17}$\,cm) is many times larger than the stellar radius ($\approx10^{12}$\,cm for a main-sequence O star).
The wind density at the boundary is given by the conservation of mass 
according to $\rho(r) = \dot{M} / 4 \pi r^2 v_\infty$, with $\dot{M}=10^{-7}$\,M$_\odot$\,yr$^{-1}$ and $v_\infty=2000$\,km\,s$^{-1}$.
Stellar rotation induces non-radial motion in the wind, but this is negligible at radius $10^{17}$\,cm.
It is, however, important for winding up the surface magnetic field of the star into a toroidal pattern, because the radial field $B_r\propto r^{-2}$ decreases much more rapidly with distance than the toroidal field, $B_\phi\propto r^{-1}$.
Stellar rotation is typically defined by the rotation velocity of the stellar surface at the equator, denoted $v_\mathrm{rot}$.

We followed the prescription of \citet{PogZanOgi04} to calculate the toroidal and radial components of the magnetic field within the boundary region based on
\begin{enumerate}
    \item the ratio of wind velocity to rotational velocity, $v_\infty/v_\mathrm{rot}$;
    \item the ratio of the distance, $d_i$, between each grid cell $i$ and the star, to the stellar radius, $d_i/R_\star$; and
    \item the latitude of the cell with respect to the stellar rotation axis.
\end{enumerate}
We assumed the rotation and magnetic axes coincide for the practical reason that on parsec scales we cannot resolve the oscillation in the equatorial current sheet that arises if the axes are misaligned.
The axis was chosen to be the +$\hat{z}$ direction.

The +$\hat{x}$ domain boundary was set to inflow with the initially uniform values of $\rho_0=10^{-23}$\,g\,cm$^{-3}$, gas pressure of $p=1.05\times10^{-11}$\,dyne\,cm$^{-2}$, and $\bm{v} = [-v_\star,0,0]$.
The gas temperature corresponding to $p=1.05\times10^{-11}$\,dyne\,cm$^{-2}$, assuming hydrogen and helium are singly ionised, is $T\approx8\,500$\,K, appropriate for photoionised ISM in the Galaxy.
The ambient magnetic field $\bm{B_0}$ was set to
\begin{enumerate}
    \item zero for HD simulations,
    \item $\bm{B_0}=4\,\mu\mathrm{G} \hat{\bm{z}}$ for 2D MHD simulations,
    \item $\bm{B_0}=[4,0,0]\,\mu\mathrm{G}$ for 3D simulations with $\bm{B_0}\cdot\bm{v_\star}=-1$, and
    \item to $\bm{B_0}=[1,4,0]\,\mu\mathrm{G}$ for 3D simulations where $\bm{B_0}$ is close to perpendicular to $\bm{v_\star}$.
\end{enumerate}

For these four cases, the plasma beta parameter is $\beta \equiv 8\pi p / \vert\bm{B_0}\vert^2 = [\infty,16.5,16.5,15.5]$, respectively.
In addition to these being the inflow gas properties, the initial domain is set to these constant values everywhere except the stellar wind inflow boundary.
The other outer domain boundaries were set to ensure outflowing gas: zero-gradient conditions if the flow direction is away from the domain, and zero velocity if an inflow tries to establish itself.

\subsection{Riemann solvers}
\label{sec:methods:solver}
In \citet{GreMacHaw19} the flux-vector splitting solver was used to calculate intercell flux; this is similar to the Harten-Lax-van Leer (HLL) Riemann solver \citep{HarLaxVan83} in that it cannot resolve CDs.
The diffusiveness was overcome by high-resolution simulations in 2D, not possible in 3D.
Similarly, \citet{GreMacKav22} used the MHD version of the HLL solver \citep[e.g.][]{MiyKus05, MigZanTze12} to ensure a robust solution for large Mach-number shocks, but this also cannot resolve the CD because the contact wave is not included in the solver.
This resulted in very smooth flows and results for predicted X-ray emission that had not converged with resolution \citep{GreMacKav22}.

To improve this, we have implemented hybrid solvers for HD and MHD.
The hybrid MHD solver uses the HLLD Riemann solver \citep{MiyKus05} for most cases, but switches to the HLL solver for cases with strong shocks and when the density contrast between the left and right states exceeds a factor of five.
The latter case is needed because most failures of the HLLD solver were found to be caused by negative density arising at the strong CD between wind and ISM, where the density contrast is a factor $10^3-10^4$.
This solver was used in \citet{MacJonBro23} to model colliding winds in binary systems.
The hybrid HD solver uses a Roe solver in conserved variables \citep{Roe81} for most solves, and the HLL solver for strong shocks and when the density contrast between left and right states exceeds a factor of ten.

\begin{table}
  \centering
  \caption{
    Simulations run for the study of bow shock properties. 
  }
  \label{tab:sims}
  \begin{tabular}{ c | c  c  c  c c }
    \hline
    Sim & Dim. & eqn. & sol. & resolution  & $\bm{B_0}\,(\mu\mathrm{G})$ \\
    \hline
    \texttt{2D-1-hd8} & 2D & HD    & 8 & $128\times 64$ & $[0,0,0]$  \\
    \texttt{2D-1-hd9} & 2D & HD    & 9 & $128\times 64$ & $[0,0,0]$  \\
    \texttt{2D-1-bx8} & 2D & MHD   & 8 & $128\times 64$ & $[4,0,0]$ \\
    \texttt{2D-1-bx7} & 2D & MHD   & 7 & $128\times 64$ & $[4,0,0]$ \\
    \texttt{3D-1-hd8} & 3D & HD    & 8 & $128^3$        & $[0,0,0]$ \\
    \texttt{3D-1-hd9} & 3D & HD    & 9 & $128^3$        & $[0,0,0]$ \\
    \texttt{3D-1-bx8} & 3D & MHD   & 8 & $128^3$        & $[4,0,0]$ \\
    \texttt{3D-1-bx7} & 3D & MHD   & 7 & $128^3$        & $[4,0,0]$ \\
    \texttt{3D-1-by7} & 3D & MHD   & 7 & $128^3$        & $[1,4,0]$ \\
    \hline
    \texttt{2D-2-hd8} & 2D & HD    & 8 & $256\times 128$ & $[0,0,0]$ \\
    \texttt{2D-2-hd9} & 2D & HD    & 9 & $256\times 128$ & $[0,0,0]$ \\
    \texttt{2D-2-bx8} & 2D & MHD   & 8 & $256\times 128$ & $[4,0,0]$ \\
    \texttt{2D-2-bx7} & 2D & MHD   & 7 & $256\times 128$ & $[4,0,0]$ \\
    \texttt{3D-2-hd8} & 3D & HD    & 8 & $256^3$         & $[0,0,0]$ \\
    \texttt{3D-2-hd9} & 3D & HD    & 9 & $256^3$         & $[0,0,0]$ \\
    \texttt{3D-2-bx8} & 3D & MHD   & 8 & $256^3$         & $[4,0,0]$ \\
    \texttt{3D-2-bx7} & 3D & MHD   & 7 & $256^3$         & $[4,0,0]$ \\
    \texttt{3D-2-by7} & 3D & MHD   & 7 & $256^3$         & $[1,4,0]$ \\
   \hline
    \texttt{2D-3-hd8} & 2D & HD    & 8 & $384\times 192$ & $[0,0,0]$ \\
    \texttt{2D-3-hd9} & 2D & HD    & 9 & $384\times 192$ & $[0,0,0]$ \\
    \texttt{2D-3-bx8} & 2D & MHD   & 8 & $384\times 192$ & $[4,0,0]$ \\
    \texttt{2D-3-bx7} & 2D & MHD   & 7 & $384\times 192$ & $[4,0,0]$ \\
    \texttt{3D-3-bx7} & 3D & MHD   & 7 & $384^3$         & $[4,0,0]$ \\
    \texttt{3D-3-by7} & 3D & MHD   & 7 & $384^3$         & $[1,4,0]$ \\
    \hline
    \texttt{2D-4-hd8} & 2D & HD    & 8 & $512\times 256$ & $[0,0,0]$ \\
    \texttt{2D-4-hd9} & 2D & HD    & 9 & $512\times 256$ & $[0,0,0]$ \\
    \texttt{2D-4-bx8} & 2D & MHD   & 8 & $512\times 256$ & $[4,0,0]$ \\
    \texttt{2D-4-bx7} & 2D & MHD   & 7 & $512\times 256$ & $[4,0,0]$ \\
    \hline
    \texttt{2D-5-hd8} & 2D & HD    & 8 & $1024\times 512$ & $[0,0,0]$ \\
    \texttt{2D-5-hd9} & 2D & HD    & 9 & $1024\times 512$ & $[0,0,0]$ \\
    \texttt{2D-5-bx8} & 2D & MHD   & 8 & $1024\times 512$ & $[4,0,0]$ \\
    \texttt{2D-5-bx7} & 2D & MHD   & 7 & $1024\times 512$ & $[4,0,0]$ \\
    \hline
  \end{tabular}
  \tablefoot{
  All simulations use the same stellar wind parameters and ISM density; only the dimensionality, equations, solver, magnetic-field orientation and resolution are varied.
  The simulations have 3 levels of refinement, focussed on the apex of the bow shock.
  Columns are, respectively, (1) simulation id, (2) dimensionality, (3) equations solved, (4) Riemann solver, (5) spatial resolution per level, and (6) ISM magnetic field orientation, where in 2D the vector components refer to $[z,R,\theta]$ and in 3D to $[x,y,z]$.
  Solver numbers correspond to: 7 = HLLD/HLL hybrid MHD solver, 8 = HLL solver (HD and MHD), 9 = hybrid Roe/HLL HD solver (see section~\ref{sec:methods:solver}).
  }
\end{table}

\subsection{Simulations}
\label{sec:methods:sims}
A set of 2D and 3D, HD and MHD simulations were performed using in-house computing resources and a grant of computing time on the HPC system \textit{Kay} from the Irish Centre for High-End Computing\footnote{\href{https://www.ichec.ie/about/infrastructure/kay}{https://www.ichec.ie/about/infrastructure/kay}}.
In total about 1.5 million core-hours were used, the majority on the highest resolution 3D simulations.

Using a nested-grid, each refinement level has the same grid shape and number of cells in each direction; only the cell diameter and domain extents change.
In 2D, simulations were performed with each refinement level containing $128\times64$, $256\times128$, $384\times192$, $512\times256$ and $1024\times512$ cells.
In 3D, resolutions of $128^3$, $256^3$ and $384^3$ cells per refinement level were achievable with the available resources.
For illustration, Fig.~\ref{fig:grid} shows a slice through a 3D HD simulation with $256^3$ zones in total per refinement level, and so each sub-domain block (bounded by the black lines) contains $64^3$ zones.
For each case in 3D, the simulations were run with HD, with MHD and $\bm{v_\star}\parallel\bm{B_0}$ (B-parallel), and with MHD and close to $\bm{v_\star}\perp\bm{B_0}$ (B-perpendicular); see above for exact values of $\bm{B_0}$.
In 2D, only the HD and B-parallel cases can be simulated because of the symmetry constraints.

The list of simulations is given in Tab.~\ref{tab:sims} with their respective properties.
In most cases, the simulations were run for at least 0.5\,Myr, but in some cases somewhat shorter when it was clear that they had reached a steady state, or when computing resources were exhausted in the case of 3D simulations.

\subsection{Post-processing of simulations}
\label{sec:methods:postprocess}
We mostly follow the methods described in \citet{MacGreMou21} and \citet{GreMacKav22} to create synthetic emission maps and integrated quantities from the simulation data as follows.

X-ray luminosities were calculated using the \textsc{pypion} python module\footnote{\href{https://github.com/greensh16/PyPion}{https://github.com/greensh16/PyPion}} to read snapshots into python, and then we used tabulated X-ray emissivities, $\eta_\mathrm{X}$, in different energy bands as a function of temperature, calculated with \textsc{Xspec} \citep{Arn96} version 12.13.1 from \textsc{heasoft} 6.32, assuming collisional ionisation equilibrium and the \textsc{apec} emission model with \citet{AspGreSau09} abundances.
The X-ray luminosity in each band was then calculated according to $L_\mathrm{X} = \sum_\mathrm{cells,i} \eta_\mathrm{X}(T_i) n_i^2 V_i$, where $n_i$  and $V_i$ are the electron number density and cell volume of cell $i$, which has temperature $T_i$.
X-ray emission maps were then generated by first loading the python arrays for $\eta_\mathrm{X}$ into the visualisation and analysis package \texttt{yt} \citep{TurSmiOis11} as a volumetric AMR dataset.
We used the \texttt{yt} \textit{ProjectionPlot} method to integrate the emissivity through the simulation box to obtain emission maps.
In all plots we considered X-rays from 0.3-10\,keV, although the emission is strongly dominated by soft X-rays and the results would be almost identical if only the range 0.3-1\,keV were plotted.
Hard X-rays (2-10\,keV) are at least two orders of magnitude fainter.

Thermal IR emission from interstellar dust grains was calculated using the \textsc{torus} radiation-hydrodynamics code \citep{HarHawAcr19} in post-processing mode to calculate the radiative equilibrium temperature of dust grains and consequent IR emission at different wavelengths, following \citet{GreMacKav22}.
The input spectrum was a \textsc{tlusty} model atmosphere of a star with effective temperature $T_\mathrm{eff}=34\,900$\,K, Solar metallicity (mass fraction of metals, $Z=0.014$), mass of 25\,M$_\odot$ and radius of 8.6\,R$_\odot$, giving a source luminosity $L=9.73\times10^4\,\mathrm{L}_\odot$ and ionising photon output $Q_0 = 3.0\times10^{48}$\,s$^{-1}$.
This corresponds approximately to a main sequence O7V-O8V star \citep{MarSchHil05}.
We assumed silicate grains with a \citet{MatRumNor77} grain-size distribution from 50\,nm to 2\,$\mu$m with a gas-to-dust ratio of 100.
A tracer variable, $y_\mathrm{w}$, was used in \textsc{pion} to distinguish between wind ($y_\mathrm{w}=1$) and ISM ($y_\mathrm{w}=0$) gas. This was used in \textsc{torus} to enforce that wind material is dust free, ISM material has the above dust properties, and mixed gas is a linear combination of the two, namely, $y_\mathrm{w} = \rho_\mathrm{w} / \rho$, where $\rho_\mathrm{w}$ is the mass density of wind material.
Additionally, we enforced that any gas with $T>10^6$\,K is dust free.
The mass fraction of dust is then given by
\begin{equation}
    f_\mathrm{D} (y_\mathrm{w},T) =
    \left\{
    \begin{array}{lr}
       0.01 (1-y_\mathrm{w})& \forall\quad T<10^6\,\mathrm{K}\\
       0                    & \forall\quad T\geq10^6\,\mathrm{K}
    \end{array}  \right\}
\end{equation}
The results are moderately sensitive to the choice of dust properties (grain-size distribution and composition, which may be non-standard because of shock processing and the EUV radiation field) but this is not crucial because we are not aiming to model any specific object in this study.
Instead we want to look at the morphological comparison between X-ray emission from hot gas and thermal dust emission from swept-up interstellar gas, for which (as we will see) the precise dust emissivity is not required.
For radiative transfer, \textsc{torus} tracks the attenuation by both gas and dust and the associated photoheating.
Nevertheless, we keep the gas temperature fixed at the value computed by \textsc{pion}, while allowing the dust to relax to its radiative equilibrium temperature.
These assumptions are appropriate because the gas density is too low for efficient collisional coupling between gas and dust temperatures.
We do not consider stochastic heating of  grains, which can raise the instantaneous temperature (and hence emissivity) of small grains exposed to EUV radiation \citep[e.g.][]{PavKirWie13}, and so our IR emission is a lower limit for the chosen grain composition and size distribution.

With these choices for dust post-processing, we are explicitly assuming wind-supported bow shocks in the strong (dynamical) coupling limit \citep{HenArt19a}, where dust and gas are tightly coupled by Coulomb interactions and the Lorentz force from the local magnetic field.
It has been suggested that under certain conditions the dust grains may decouple from the gas via relative drift motions driven by radiation pressure on dust grains \citep{Dra11, OchCoxKri14, AkiKirPav15} or a large gyroradius for large dust grains \citep{KatAleGva18}.
This can lead to IR morphology that is not representative of the gas density because of the dynamical decoupling, although a detailed treatment of all the relevant forces by \citet{HenArt19b} shows that radiation-driven dust waves may only form for O stars moving with $v_\star\approx150-300$\,km\,s$^{-1}$, unless the stars have exceptionally weak winds.
Our assumptions are therefore likely to be valid, except that the potentially large gyroradius for the biggest grains could make the CD less sharp than we predict.
Dust grains may also survive for some time if entrained in the hot gas of the shocked stellar-wind \citep[e.g.][]{EveChu10} before being destroyed by collisions.
This again would tend to make the CD more diffuse in mid-IR emission than our predictions.

\begin{figure*}
	\includegraphics[width=\textwidth]{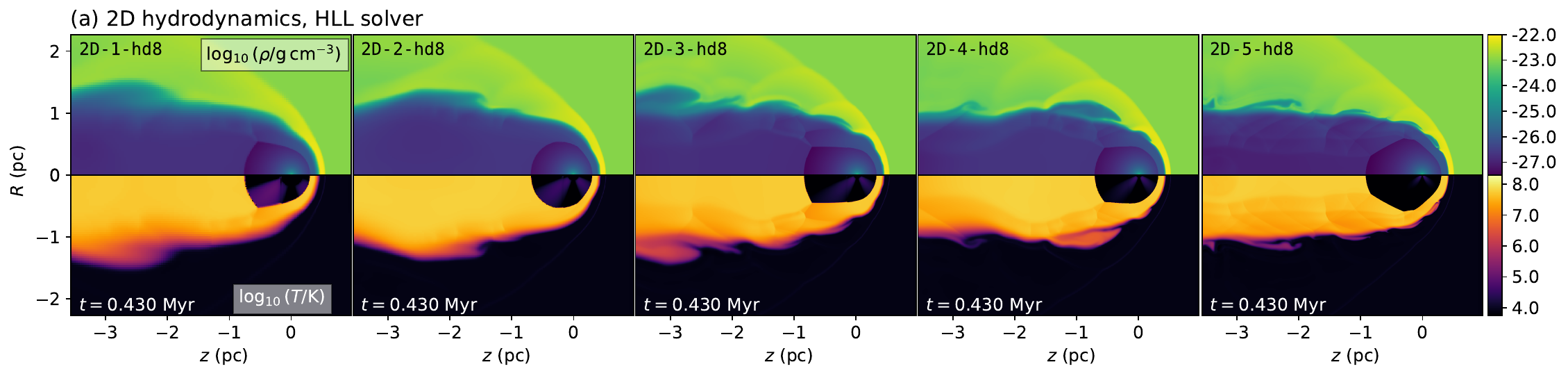}
	\includegraphics[width=\textwidth]{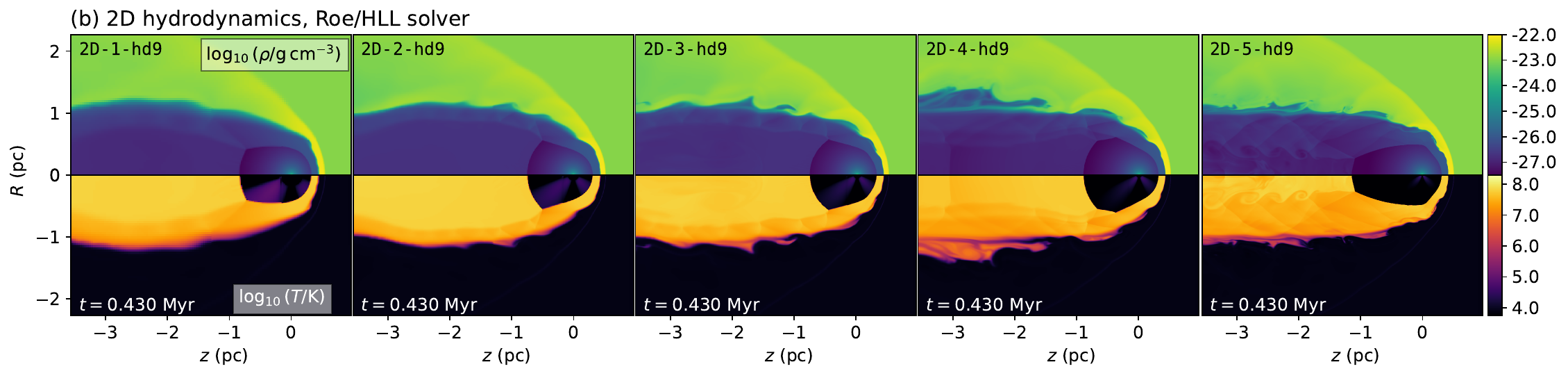}
	\includegraphics[width=\textwidth]{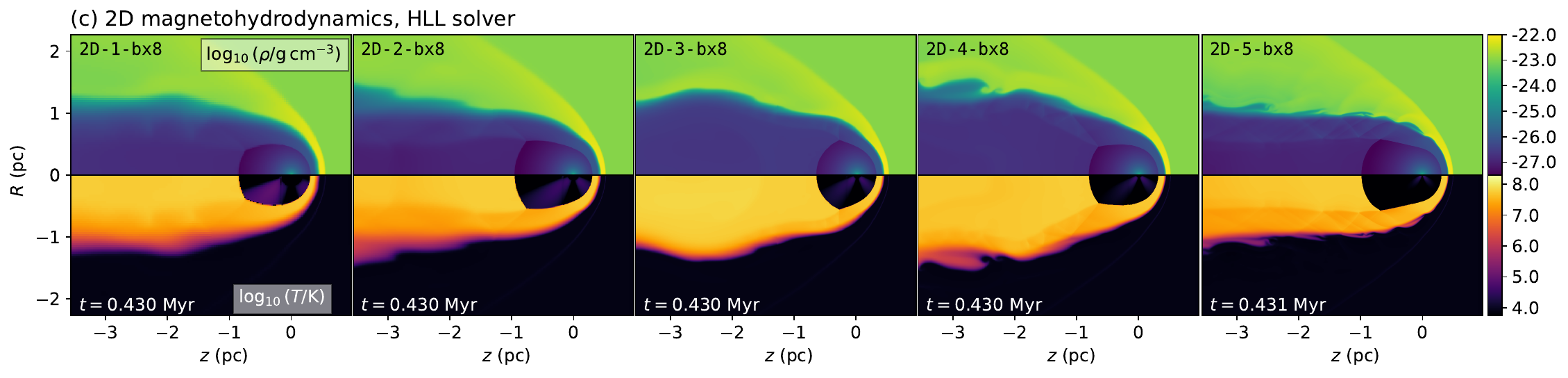}
	\includegraphics[width=\textwidth]{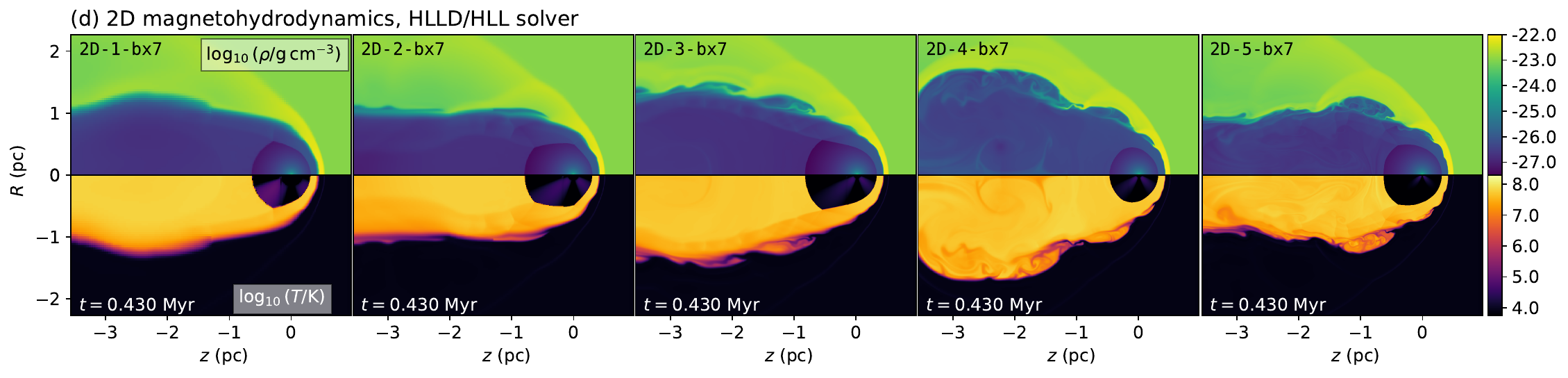}
    \caption{Gas density (temperature) in upper (lower) half-plane in 2D simulations of the bow shock at different resolution using different solvers (simulation id indicated on the upper panel of each plot).
    The top row (a) is HD simulations with the HLL solver, 2nd row (b) is HD simulations with the hybrid Roe/HLL solver, 3rd row (c) is MHD simulations using the HLL solver and 4th row (d) is MHD simulations using the hybrid HLLD/HLL solver.
    All snapshots are taken at $t=0.43$\,Myr.
    }
    \label{fig:2dresolution-sametime}
\end{figure*}

\begin{figure}
  \includegraphics[width=\columnwidth,trim={0cm 1.3cm 0 0cm}, clip]{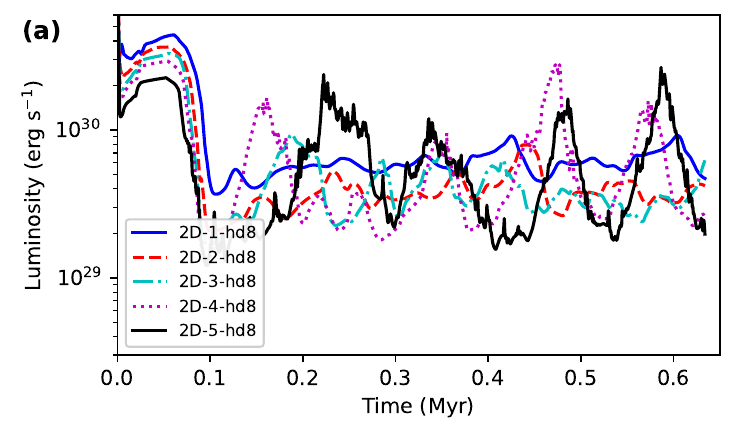}
  \includegraphics[width=\columnwidth,trim={0cm 1.3cm 0 0cm}, clip]{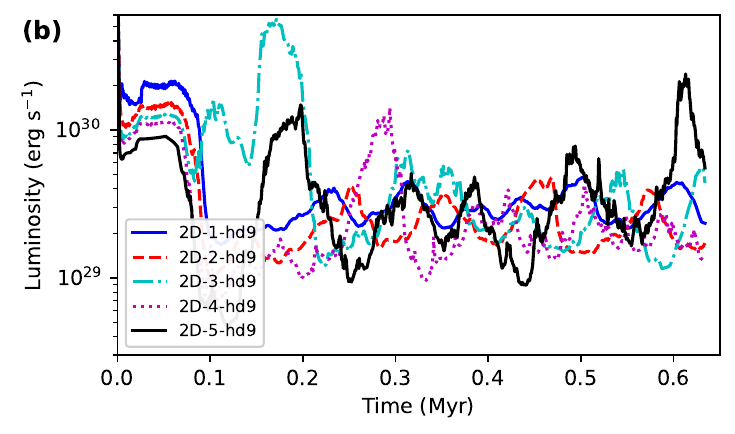}
  \includegraphics[width=\columnwidth,trim={0cm 1.3cm 0 0cm}, clip]{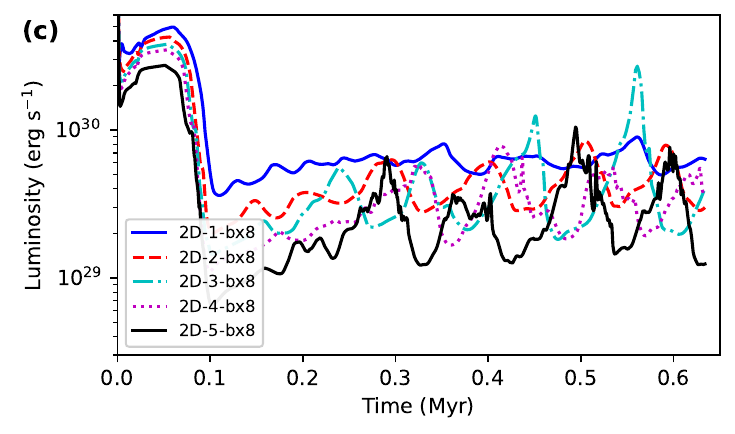}
  \includegraphics[width=\columnwidth]{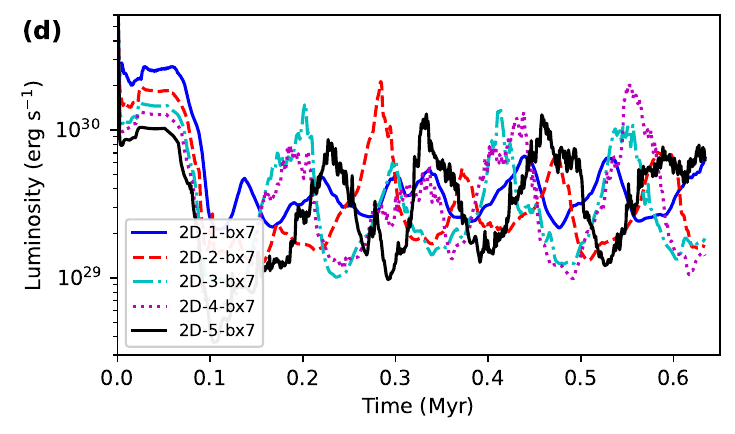}
  \caption{Thermal X-ray luminosity in 0.3-10\,keV band as a function of time for 2D HD and MHD simulations at different resolutions using the HLL and hybrid solvers.
    From top to bottom: (a) HD+HLL, (b) HD+Roe/HLL, (c) MHD+HLL, (d) MHD+HLLD/HLL.}
  \label{fig:2dsolver-xray}
\end{figure}

\section{Results} \label{sec:results}

A much larger range of resolutions and parameters can be explored with 2D simulations, with the notable exception of the orientation of the interstellar magnetic field which is constrained to be parallel to the symmetry axis.
We first discuss the 2D results, because these then informed which 3D simulations were run.
Fig.~\ref{fig:2dresolution-sametime} shows snapshots from all 2D simulations in Tab.~\ref{tab:sims} at $t=0.43$\,Myr, with log of gas density shown on the upper half-plane and log of gas temperature on the lower half-plane.

The star is located at the origin and is surrounded by a low-density and cold region of freely expanding stellar wind. The temperature structure in this region is a numerical artefact arising because the thermal pressure is negligible compared with the kinetic energy of the wind, and has no effect on the solution.
This wind interacts with a uniform ISM flowing past with relative velocity $v_\star$ from right to left.
The apex of the bow shock, in the upstream at $z\approx0.5$\,pc, has the highest density gas because here the outer shock has the largest Mach number and compression factor (the shock is approximately isothermal).
The wind termination shock is adiabatic because of the low density and large post-shock temperature, and a hot plasma with $T\approx6\times10^7$\,K expands from the apex downstream along the pressure gradient.
The ISM flow speed is $\approx v_\star=30$\,km\,s$^{-1}$, whereas the hot gas sound speed is $c_s\sim1000$\,km\,s$^{-1}$, and so a strong shear layer forms along the wind-ISM CD, with associated KH instability.

The general trends are immediately apparent: the HLL solver (rows a and c) has weak dynamical mixing but strong numerical diffusion at the CD; the hybrid solvers (rows b and d) resolve dynamical instabilities at the CD much better at a given numerical resolution and the dynamical mixing is progressively better developed as resolution increases.
This is true for both HD and MHD simulations.

\subsection{Choice of Riemann solver} \label{sec:res:solver}

Looking at low-resolution HD and MHD simulation in Fig.~\ref{fig:2dresolution-sametime}, the simulations using the diffusive HLL solver (rows (a) and (c): \texttt{2D-1-hd8}, \texttt{2D-2-hd8}, \texttt{2D-1-bx8}, \texttt{2D-2-bx8}) show a relatively smooth contact discontinuity with weak and poorly resolved waves raised on the interface, similar to the 3D MHD simulations of the bow shock of $\zeta$ Oph by \citet{GreMacKav22}.
In contrast, using the hybrid solver (rows (b) and (d): \texttt{2D-1-hd9}, \texttt{2D-2-hd9}, \texttt{2D-1-bx7}, \texttt{2D-2-bx7}) shows a sharper interface at the CD, more well-developed waves forming on the interface, and better-resolved entrainment of ISM gas into the hot wake behind the star.
With the HLL solver, the mixing/entrainment seems to be mainly diffusive, whereas with the hybrid solvers it is driven by the unstable flow dynamics at the interface.

The features in the wake behind the star are somewhat intermittent, and a given snapshot may or may not show well-developed KH instability, but one may always look at the CD in the upstream direction to see how well the KH instability is resolved.
With the hybrid solvers one can see some deformations forming near the apex of the bow shock, whereas these are difficult to see in the upper panel with the HLL solver at low resolutions (1 and 2).

At higher resolution, we may examine simulations \texttt{2D-5-hd8}, \texttt{2D-5-hd9}, \texttt{2D-5-bx8} and \texttt{2D-5-bx7}, the right-most column of Fig.~\ref{fig:2dresolution-sametime}.
Here, all simulations can well resolve the development of KH instability at multiple scales, showing interaction of small- and large-scale eddies in the non-linear phase of the instability.
Nevertheless, the CD in the upstream part of the flow remains relatively smooth with the HLL solver because, for the numerically resolved modes, the growth timescale is sufficiently long that the waves are not apparent.
With the hybrid solvers, well-developed KH rolls are already apparent in the upstream part of the CD.

The X-ray luminosity of all 2D simulations in the 0.3-10\,keV band is plotted as a function of time in Fig.~\ref{fig:2dsolver-xray}, where the top panel (a) shows HD+HLL, second panel (b) shows HD+Roe/HLL, third panel (c) shows MHD+HLL solver, and the bottom panel (d) shows MHD+HLLD/HLL solver.
Each panel shows all simulation resolutions run for that combination of equations plus solver.
The main trend in all panels is increasing time-variability with increasing spatial resolution, with a dominant variability timescale of about 0.1-0.12\,Myr in most simulations.
Simulations using the hybrid solvers (panels b and d) have stronger variability at a given resolution than those with the HLL solver (panels a and c), because X-ray variability is driven by mixing of wind and ISM gas at the CD, producing gas with intermediate density and temperature which emits most strongly in thermal X-rays.
The highest-resolution simulations also have variability on multiple timescales, reflecting the multiple scales of KH vortices present on the domain at any given time.

The initially strong emission followed by a steep drop at $t<0.12$\,Myr is a result of the initial conditions where a strongly over-pressurised bubble expands to its equilibrium size and becomes distorted by stellar motion.
This timescale and the overall variability timescale are similar to the dynamical timescale of the simulation, $\tau_\mathrm{box} = L_\mathrm{box}/v_\star \approx0.15$\,Myr, and has been seen in previous work \citep{GreMacHaw19}.
If only the downstream part of the domain ($x_\star - x_\mathrm{min} = 1.1\times10^{19}$\,cm) is considered instead of the full box, the advection timescale is $\tau_\mathrm{ad} = 0.116$\,Myr, which seems to match the variability timescale very well.
This is significantly longer than the dynamical timescale of the bow shock, $\tau_\mathrm{bs} = R_\mathrm{so} / v_\star \approx 0.01$\,Myr.
Here, the standoff distance of the bow shock, $R_0 = \sqrt{ \dot{M} v_\infty / 4\pi \rho_0 (c_s^2 + v_A^2 + v_\star^2)}$, where $c_s$ is the sound speed of the ISM and $v_A$ the Alfv\'en speed, .
The similarity of the variability timescale and $\tau_\mathrm{ad}$ suggests that the X-ray variation is initiated by discrete regions of enhanced mixing of wind and ISM, that are then advected downstream and leave the domain.
Note that the time variation at different numerical resolutions are not in phase with each other, so the snapshots plotted in Fig.~\ref{fig:2dresolution-sametime} are in some cases near the maximum and in other cases near the minimum of diffuse X-ray luminosity.

It is clear that the details of the flow have not converged with resolution, and will not converge because of the nature of instabilities seeded by discretisation errors on the grid scale.
Nevertheless, the X-ray emission varies within a range that appears to have converged, at a level $L_\mathrm{X}\approx 10^{29}-2\times10^{30}$ erg\,s$^{-1}$.
The development of KH instability is sufficiently well-resolved that the time-averaged properties of the gas at the wind-ISM interface are no longer resolution dependent, at least when using the hybrid solvers.
For the HLL solver, Fig.~\ref{fig:2dsolver-xray} shows that the time-averaged X-ray luminosity decreases as resolution increases, because the diffusivity of the solver is still contributing to producing the X-ray emitting layer \citep[cf.][]{TanOhGro21}.

\subsection{Comparison of HD and MHD simulations}
Figs.~\ref{fig:2dresolution-sametime}(a) and (c) show HD and MHD simulations, respectively, both using the HLL solver.
The results are very similar in both cases, only that the KH instability is slightly less well-developed in the MHD case.
This could be because the MHD HLL solver may be more diffusive than the HD HLL solver, or it could be a physical effect that the magnetic field lines draped along the CD are inhibiting development of KH instability.
Figs.~\ref{fig:2dresolution-sametime}(b) and (d) show the same comparison, but using the hybrid solvers that can resolve the CD, and in this case there is no clear difference between the two series of plots.
It would be difficult to say which calculation is HD and which is MHD when only looking at the density and temperature fields.
This largely reflects the initial conditions, in that neither the stellar wind nor the ISM are magnetically dominated, and so the magnetic field does not have a strong impact on the dynamics of the bow shock.

The development of KH instability, and associated entrainment of ISM gas into the hot, shocked, stellar wind, does not seem to be much affected by the magnetic field (for simulations with the hybrid solvers).
Similarly, the range of X-ray luminosities achieved in the HD and MHD simulations with the hybrid solvers in Fig.~\ref{fig:2dsolver-xray} appears to be the same; if anything the MHD simulations show stronger time-variation.
We may conclude that the presence or absence of magnetic fields does not appear to play a significant role in the dynamical evolution of the 2D bow-shock simulations, at least for the wind and ISM parameters considered for this study.
Note, however, that in this 2D configuration where the ISM magnetic field is parallel to the shock normal-vector at the apex of the bow shock, we expect the MHD and HD dynamics to be similar because the magnetic field is not compressed in the shock.
In 3D with different field configurations there are noticeable differences.

\begin{figure}
	\includegraphics[width=\columnwidth]{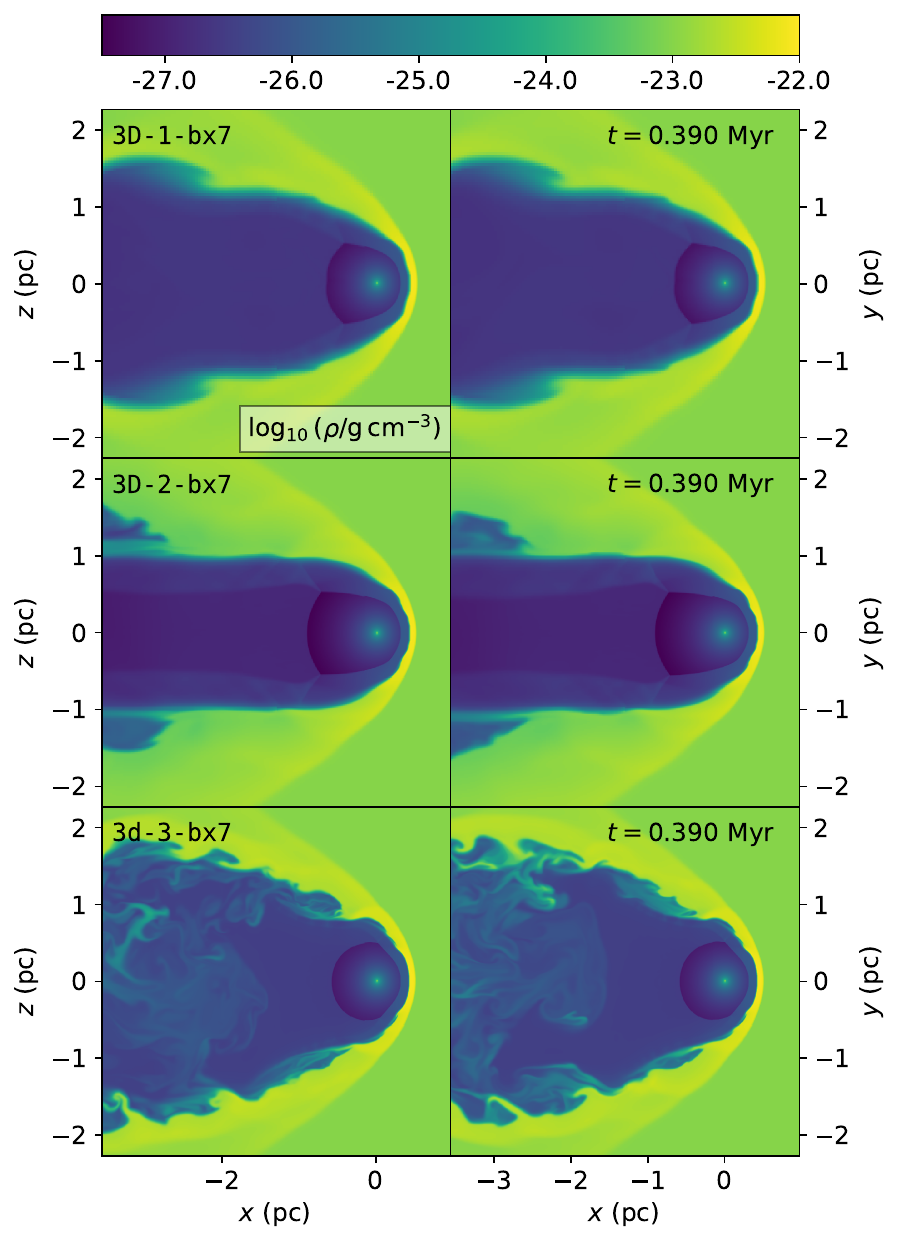}
    \caption{Slices in the planes $y=0$ (left) and $z=0$ (right) of gas density on a logarithmic density scale for simulations \texttt{3D-1-bx7} (top), \texttt{3D-2-bx7} (middle) and \texttt{3D-3-bx7} (bottom), ordered with increasing resolution from top to bottom.  These simulations have the ISM magnetic field in configuration 1, parallel to the space velocity of the star through the ISM.
    All snapshots are taken at $t=0.39$\,Myr.
    }
    \label{fig:3d-mhd-bx-res-sametime}
\end{figure}

\begin{figure}
	\includegraphics[width=\columnwidth]{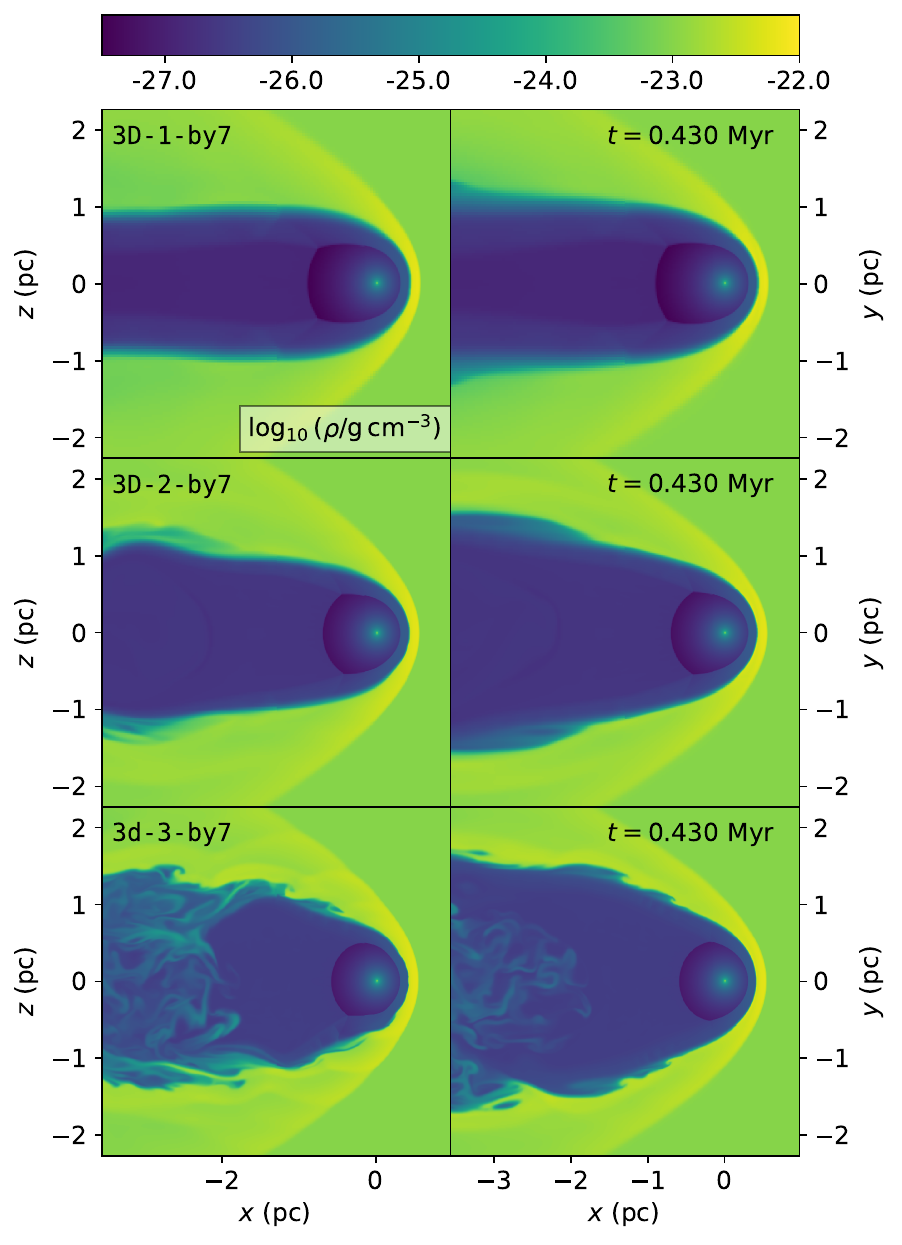}
    \caption{Slices in the planes $y=0$ (left) and $z=0$ (right) of gas density on a logarithmic density scale for simulations \texttt{3D-1-by7} (top), \texttt{3D-2-by7} (middle) and \texttt{3D-3-by7} (bottom), ordered with increasing resolution from top to bottom.  These simulations have the ISM magnetic field in configuration 2, almost perpendicular to the space velocity of the star through the ISM.
    All snapshots are taken at $t=0.43$\,Myr.
    }
    \label{fig:3d-mhd-by-res-sametime}
\end{figure}

\begin{figure}[ht]
	\centering
	\includegraphics[width=0.48\textwidth]{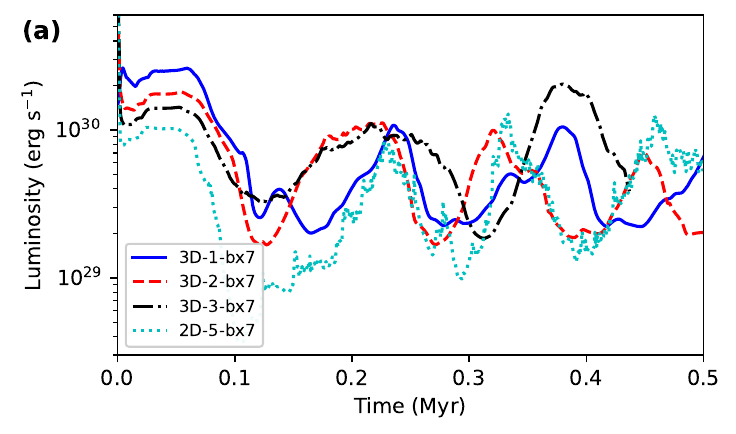}
	\includegraphics[width=0.48\textwidth]{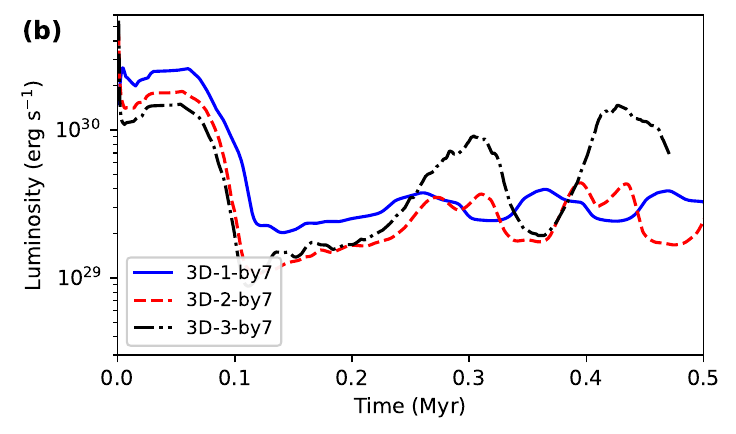}
    \caption{X-ray luminosity as a function of time, of the hot plasma in (a) the bow-shock simulations \texttt{3D-1-bx7}, \texttt{3D-2-bx7}, \texttt{3D-3-bx7}, and (b) \texttt{3D-1-by7}, \texttt{3D-2-by7}, \texttt{3D-3-by7}.
    For comparison, results for the high-resolution 2D simulation \texttt{2D-5-bx7} are also plotted in the upper panel.
    }
    \label{fig:xray-lum-3d-mhd}
\end{figure}

\begin{figure*}[ht]
	\centering
	\includegraphics[width=0.8\textwidth]{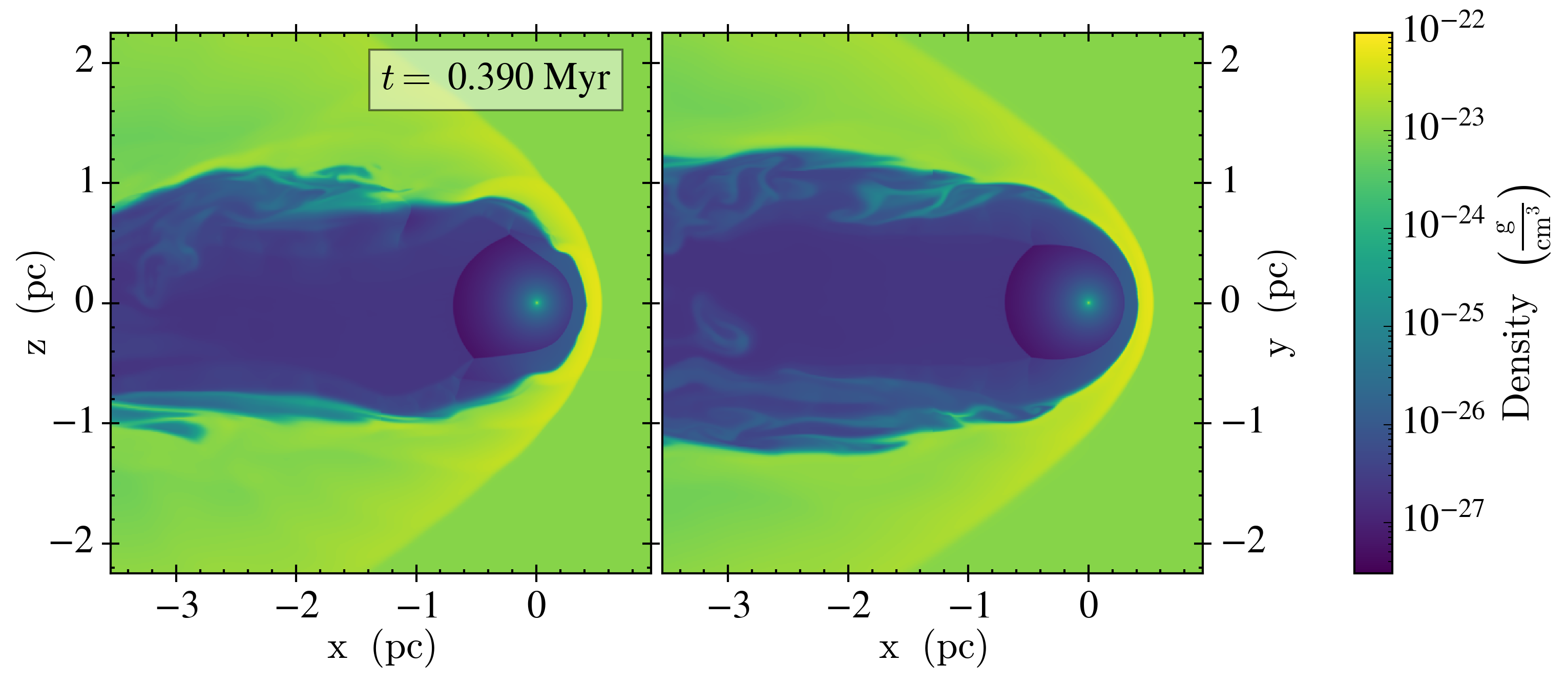}
	\includegraphics[width=0.8\textwidth]{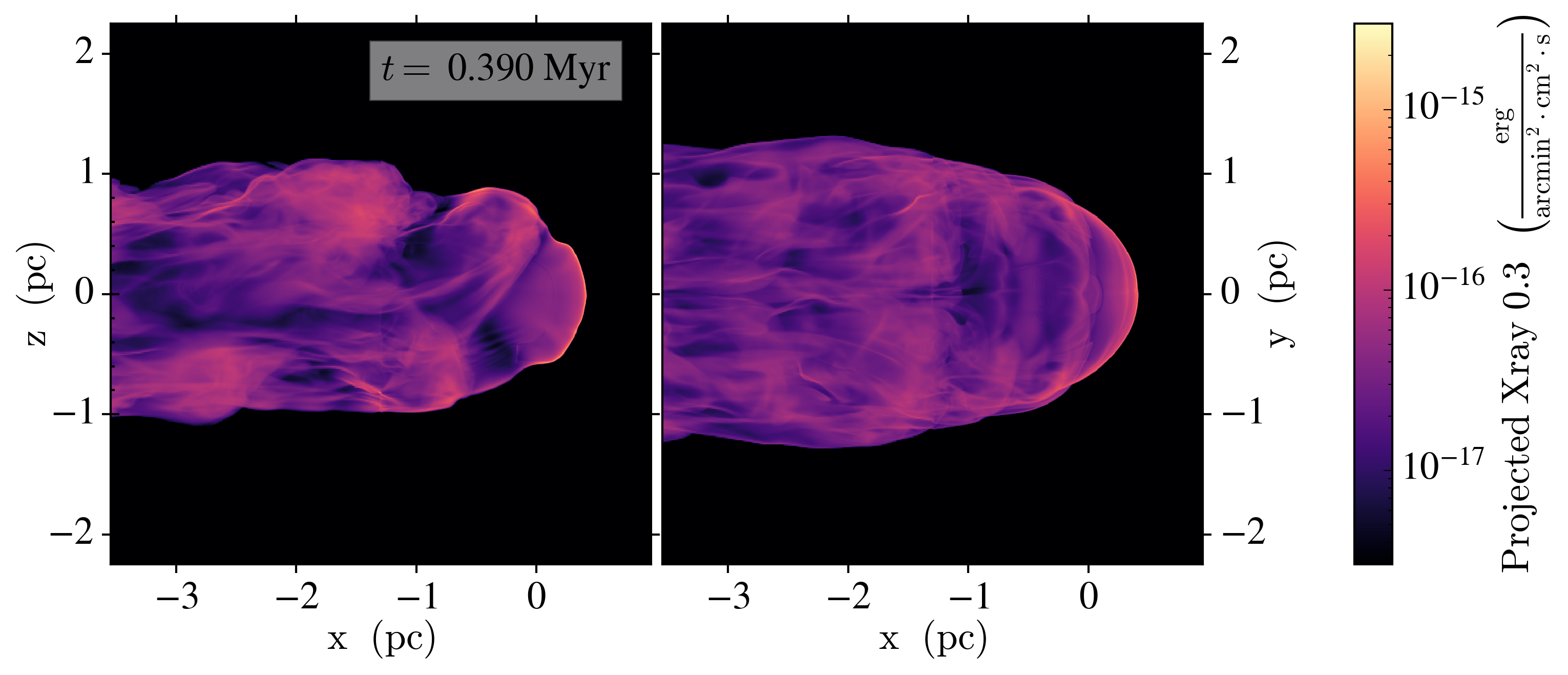}
	\includegraphics[width=0.8\textwidth]{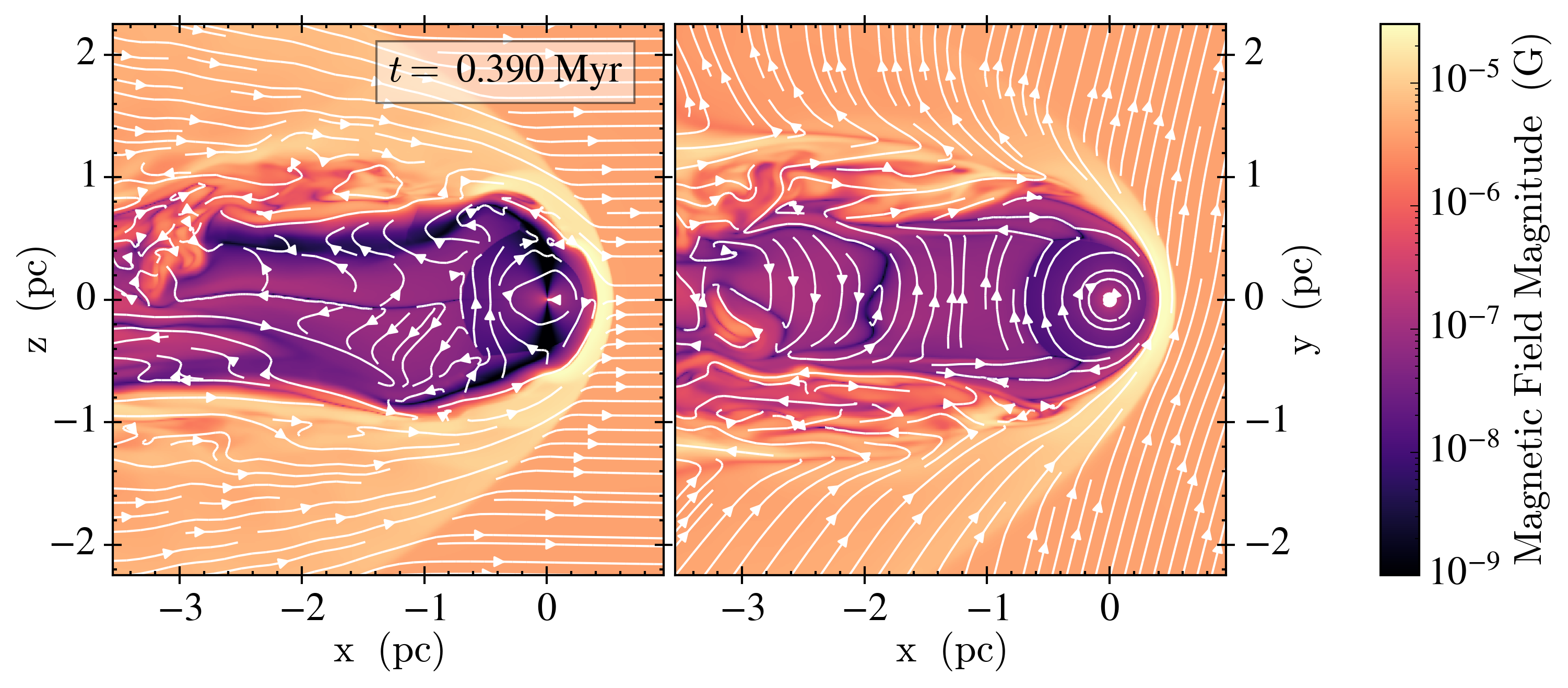}
    \caption{For simulation \texttt{3D-3-by7} at simulation time 0.39\,Myr, from top to bottom:
        slices showing gas density in the planes $y=0$ (left) and $z=0$ (right);
        projected thermal X-ray intensity in the 0.3-2\,keV band along the $\hat{y}$ (left) and $\hat{z}$ axes (right); and
        slices of magnetic field strength on a logarithmic colour scale with field orientation shown as streamlines in the planes $y=0$ (left) and $z=0$ (right).
        This simulation has the ISM magnetic field in configuration 2, almost perpendicular to the space velocity of the star through the ISM and with no $\hat{z}$ component.}
    \label{fig:3d-mhd-by-comp1}
\end{figure*}

\begin{figure*}[ht]
	\centering
	\includegraphics[width=0.8\textwidth]{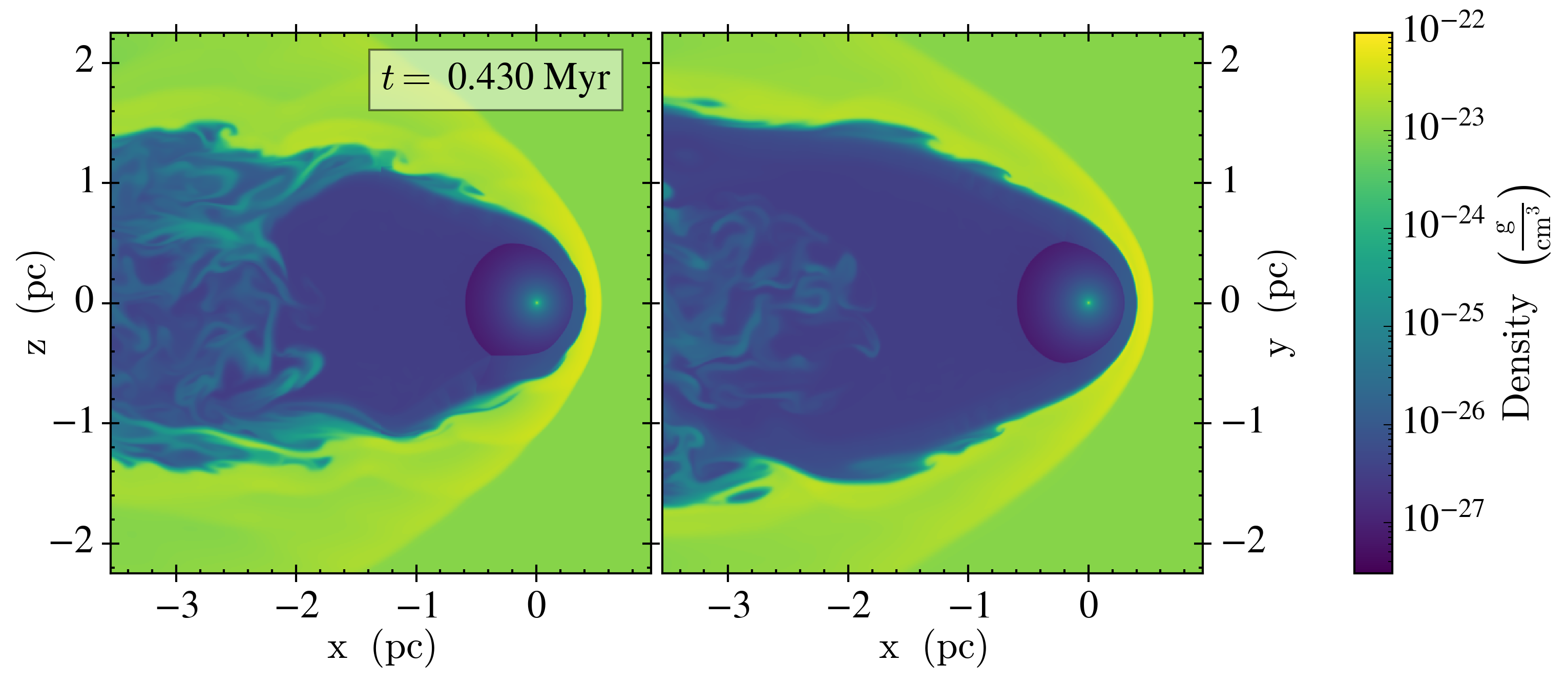}
	\includegraphics[width=0.8\textwidth]{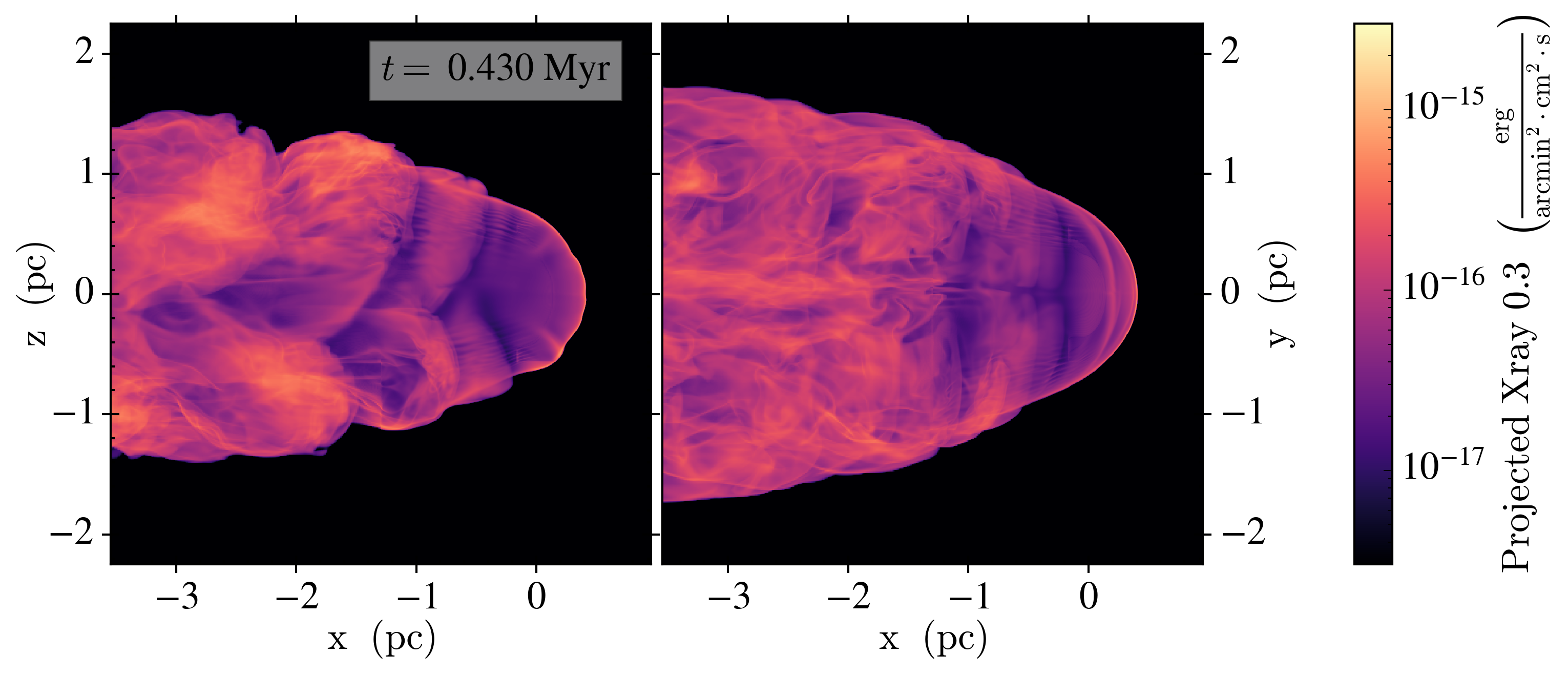}
	\includegraphics[width=0.8\textwidth]{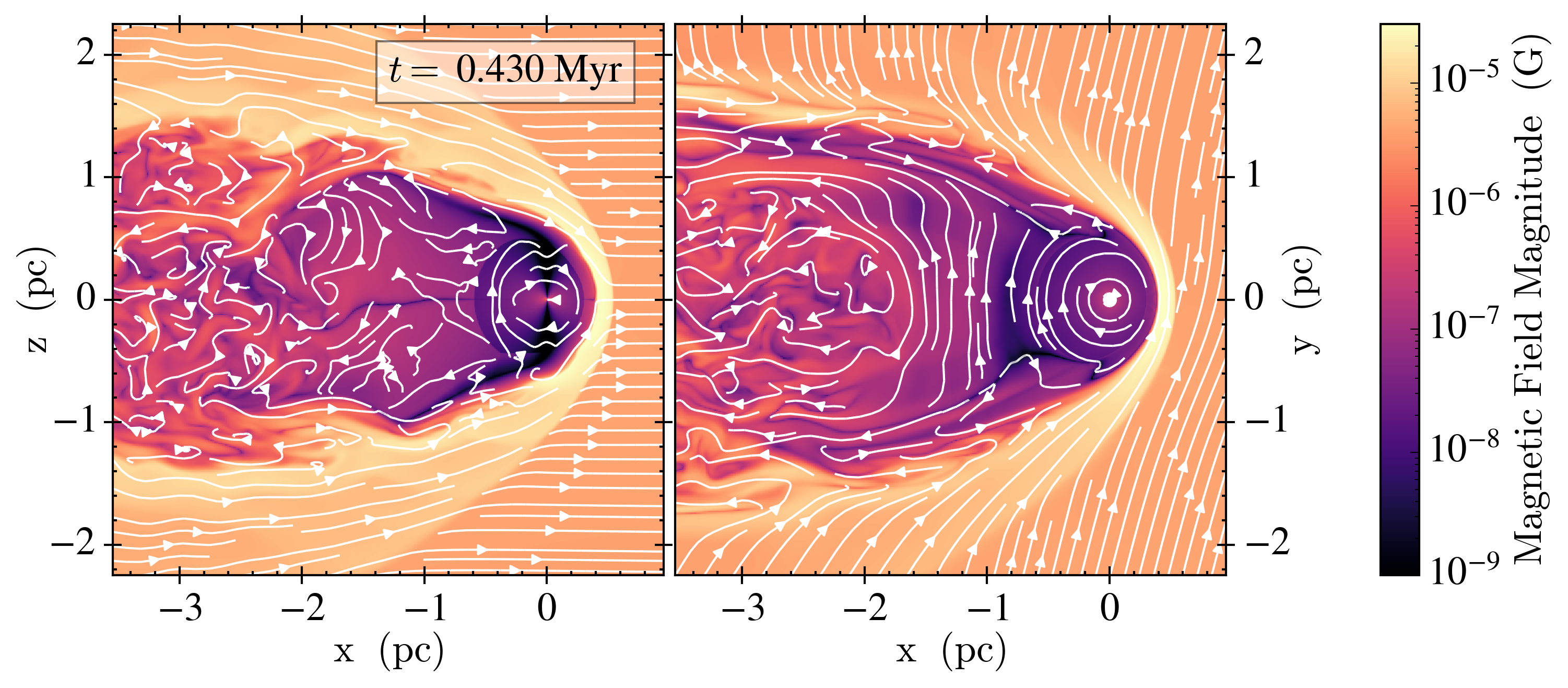}
    \caption{Same as Fig.~\ref{fig:3d-mhd-by-comp1} but at a later time of 0.43\,Myr, close to the time of maximum X-ray luminosity.}
    \label{fig:3d-mhd-by-comp2}
\end{figure*}

\begin{figure}[ht]
	\includegraphics[width=0.48\textwidth]{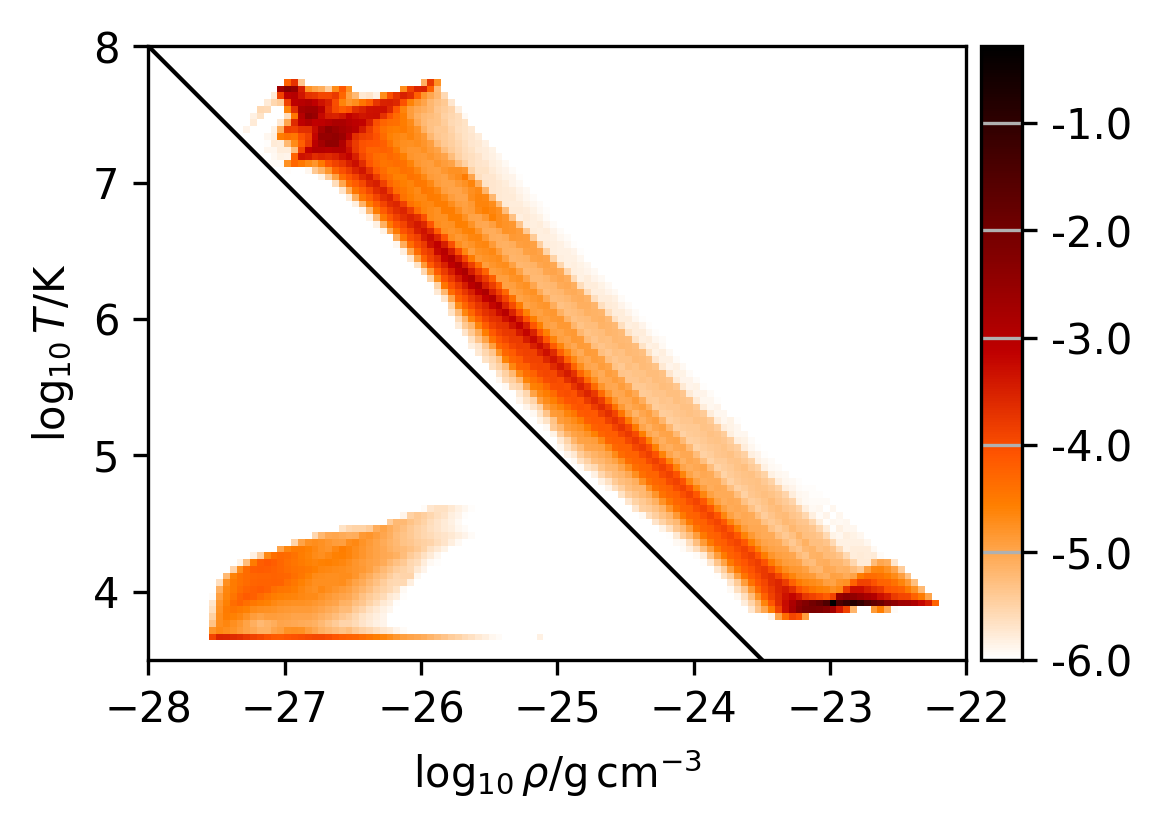}
	\includegraphics[width=0.48\textwidth]{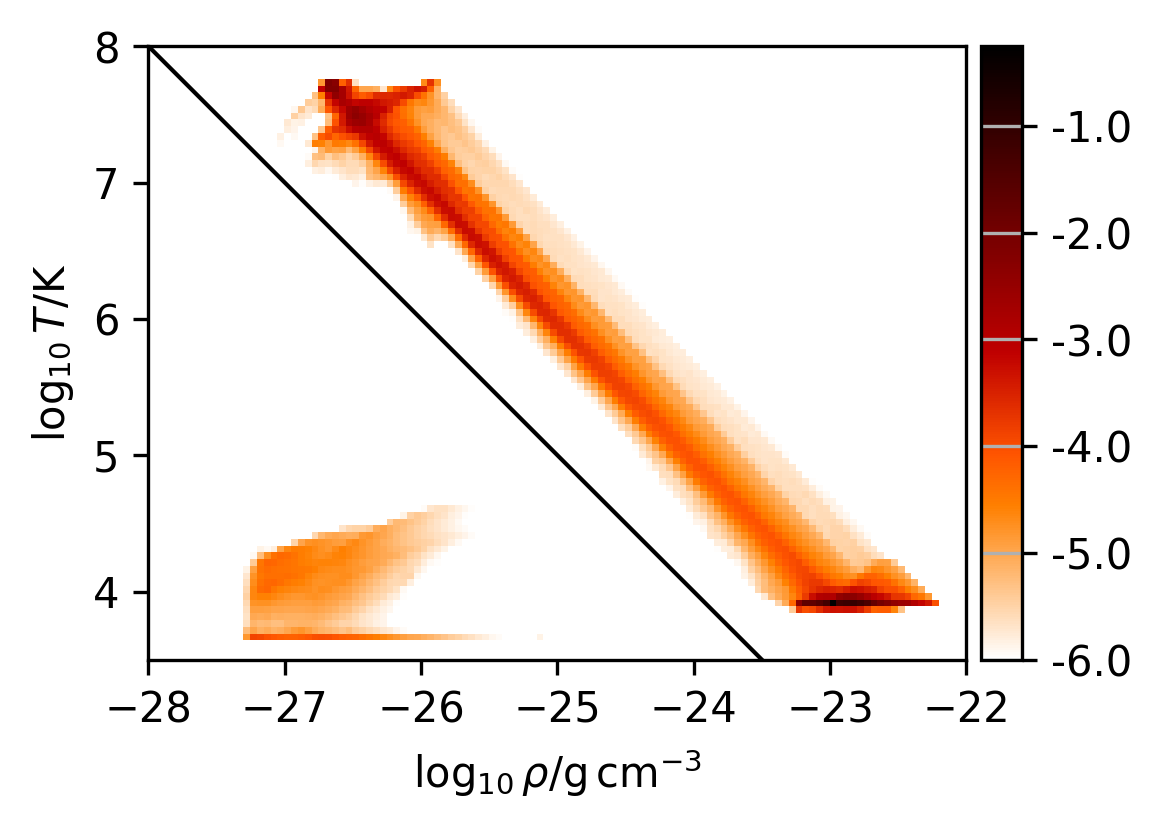}
	\includegraphics[width=0.48\textwidth]{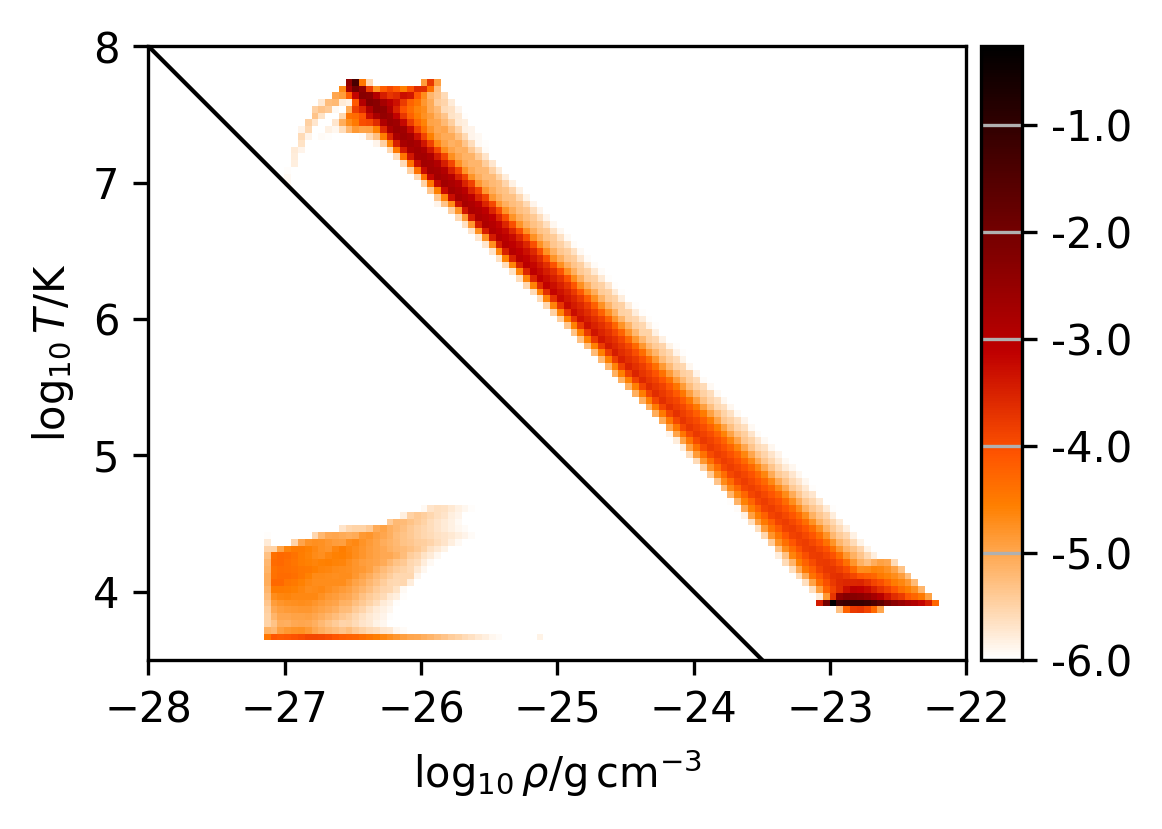}
    \caption{Volume fraction of gas in the $\rho-T$ plane for simulation \texttt{3D-3-by7} at times  $t=0.35$\,Myr (top), 0.39\,Myr (middle) and 0.43\,Myr (bottom).
    The colour scale shows $\log_{10}$ of the fraction of the simulation volume within the density and temperature range enclosed by each pixel.
    These simulations have the ISM magnetic field in configuration 2, almost perpendicular to the space velocity of the star through the ISM.
    The black line shows a line of constant pressure, $p/k_\mathrm{B} \approx 10^4$\,cm$^{-3}$\,K. }
    \label{fig:phase-3d-mhd-by}
\end{figure}

\begin{figure}[ht]
	\includegraphics[width=0.48\textwidth]{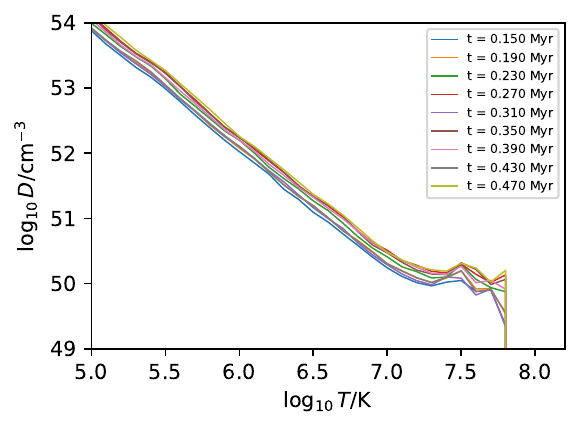}
	\includegraphics[width=0.48\textwidth]{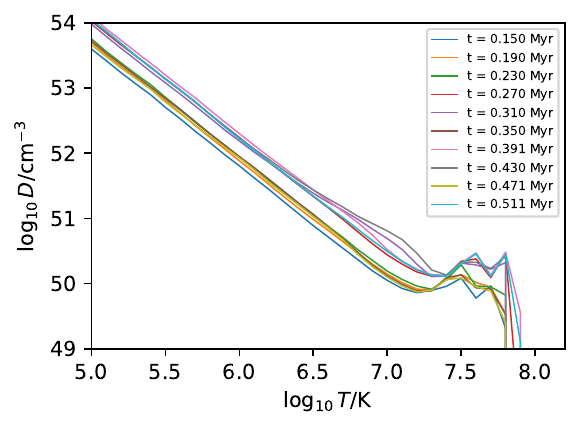}
	\includegraphics[width=0.48\textwidth]{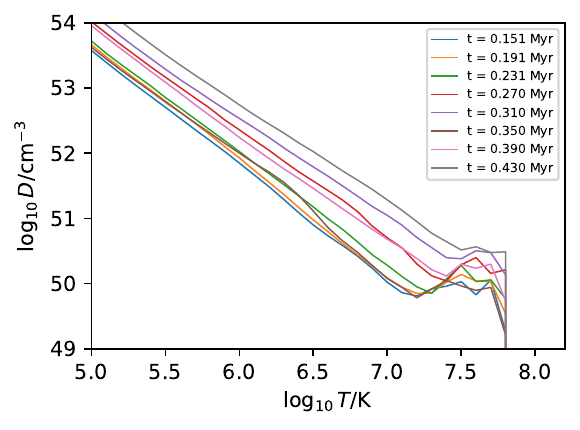}
    \caption{DEM plots at selected times as indicated for simulations \texttt{3D-1-by7} (top), \texttt{3D-2-by7} (middle) and \texttt{3D-3-by7} (bottom), with increasing resolution from top to bottom.  These simulations have the ISM magnetic field in configuration 2, almost perpendicular to the space velocity of the star through the ISM.}
    \label{fig:dem-3d-mhd-by}
\end{figure}

\begin{figure*}[ht]
	\centering
	\includegraphics[width=\textwidth]{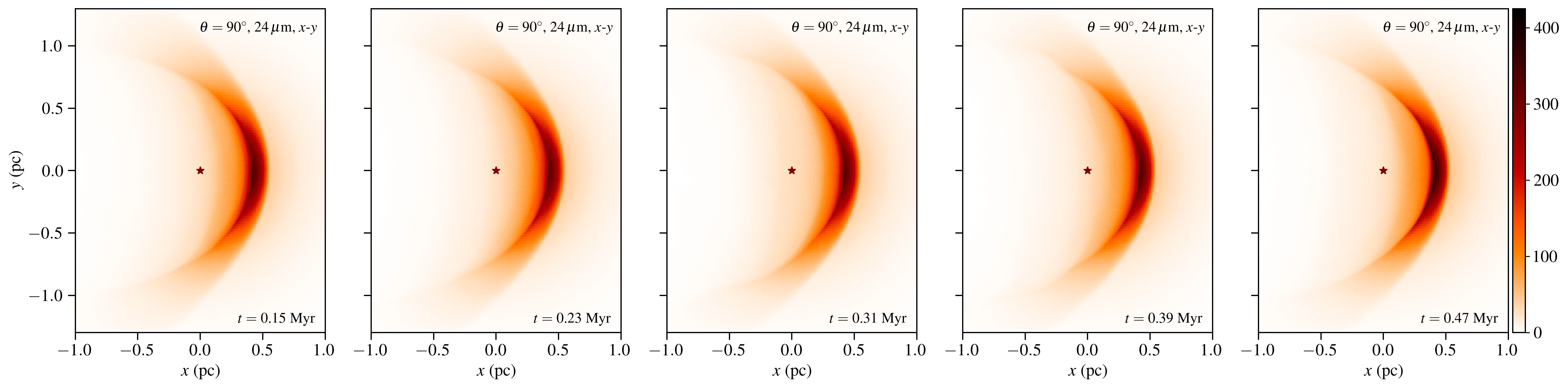}
	\includegraphics[width=\textwidth]{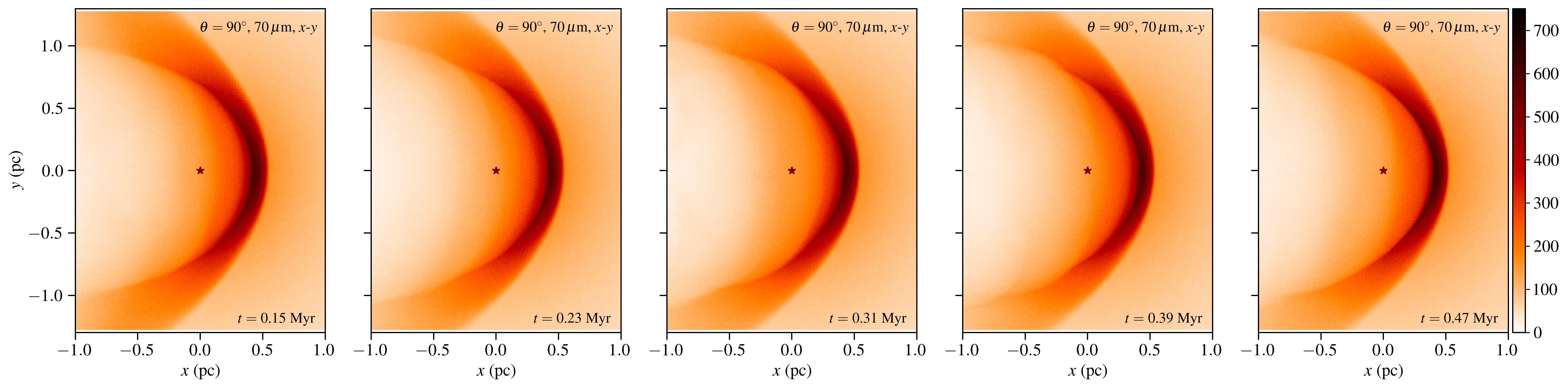}
	\includegraphics[width=\textwidth]{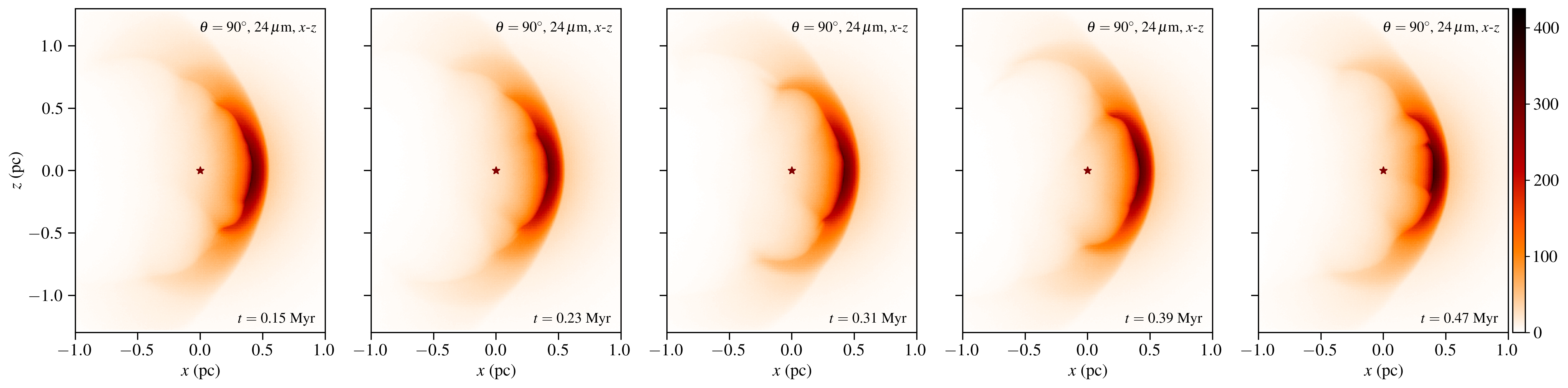}
	\includegraphics[width=\textwidth]{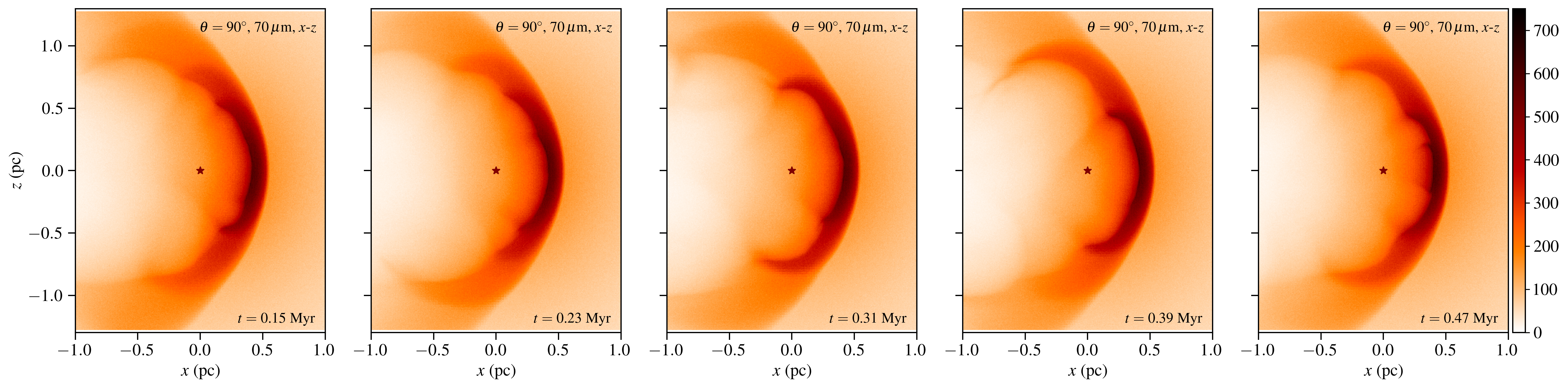}
    \caption{Infrared dust emission from a time series of snapshots for \texttt{3D-3-by7} showing, from top to bottom, respectively, 24\,$\mu$m emission for projection along $\hat{z}$, 70\,$\mu$m emission for projection along $\hat{z}$, 24\,$\mu$m emission for projection along $\hat{y}$ and 70\,$\mu$m emission for projection along $\hat{y}$.
    The linear colour scale is in units of MJy\,ster$^{-1}$, and note that the two different observation wavelengths have different colour scales.
    Snapshots are shown from left to right ever 0.08\,Myr from 0.15\,Myr to 0.47\,Myr.
    The star symbol shows the location of the star at the origin.}
    \label{fig:3d-mhd-ir}
\end{figure*}

\begin{figure*}[ht]
	\centering
	\includegraphics[height=6.5cm,trim={1cm 0cm 4cm 0cm}]{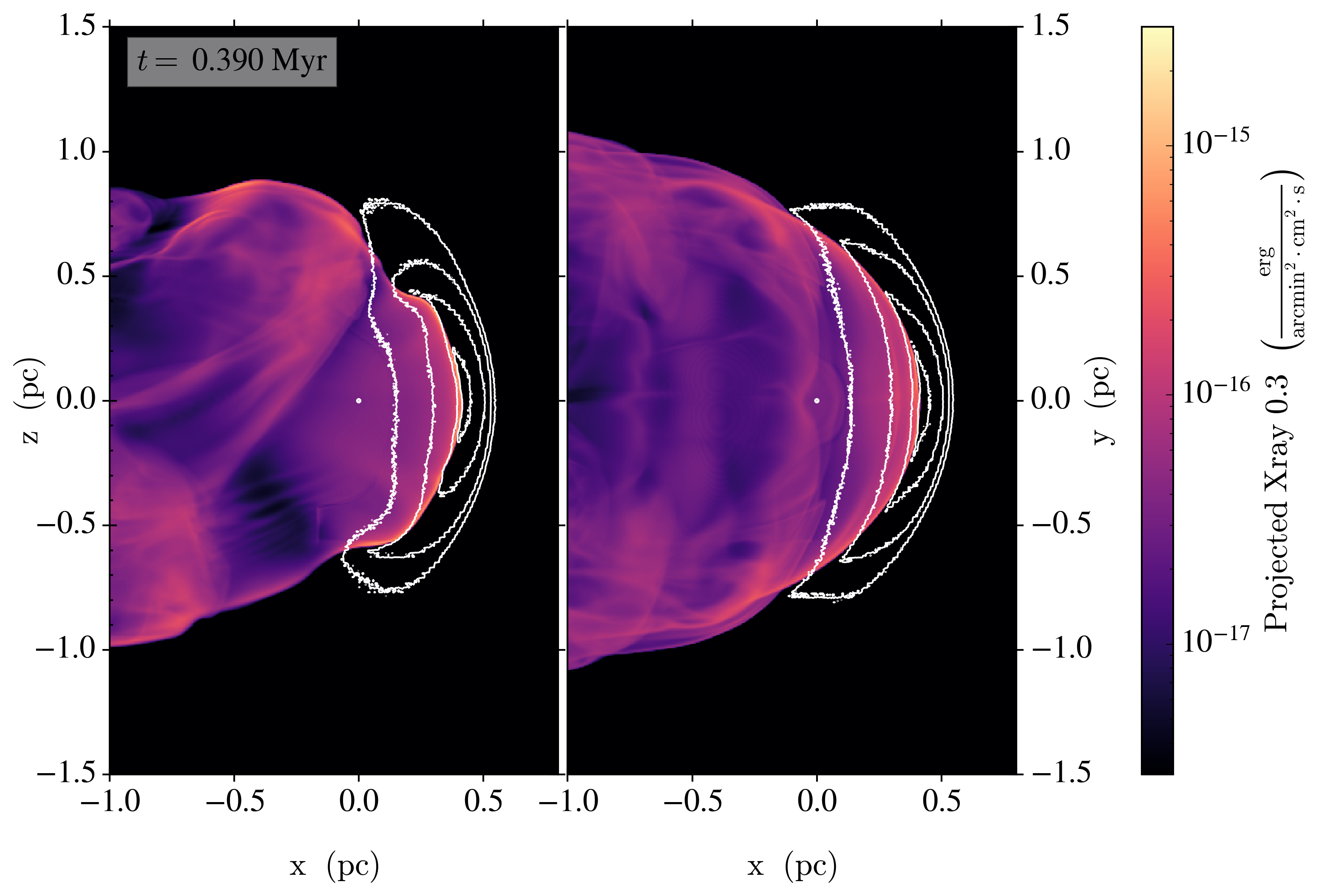}
	\includegraphics[height=6.5cm,trim={0cm 0cm 0cm 0cm}]{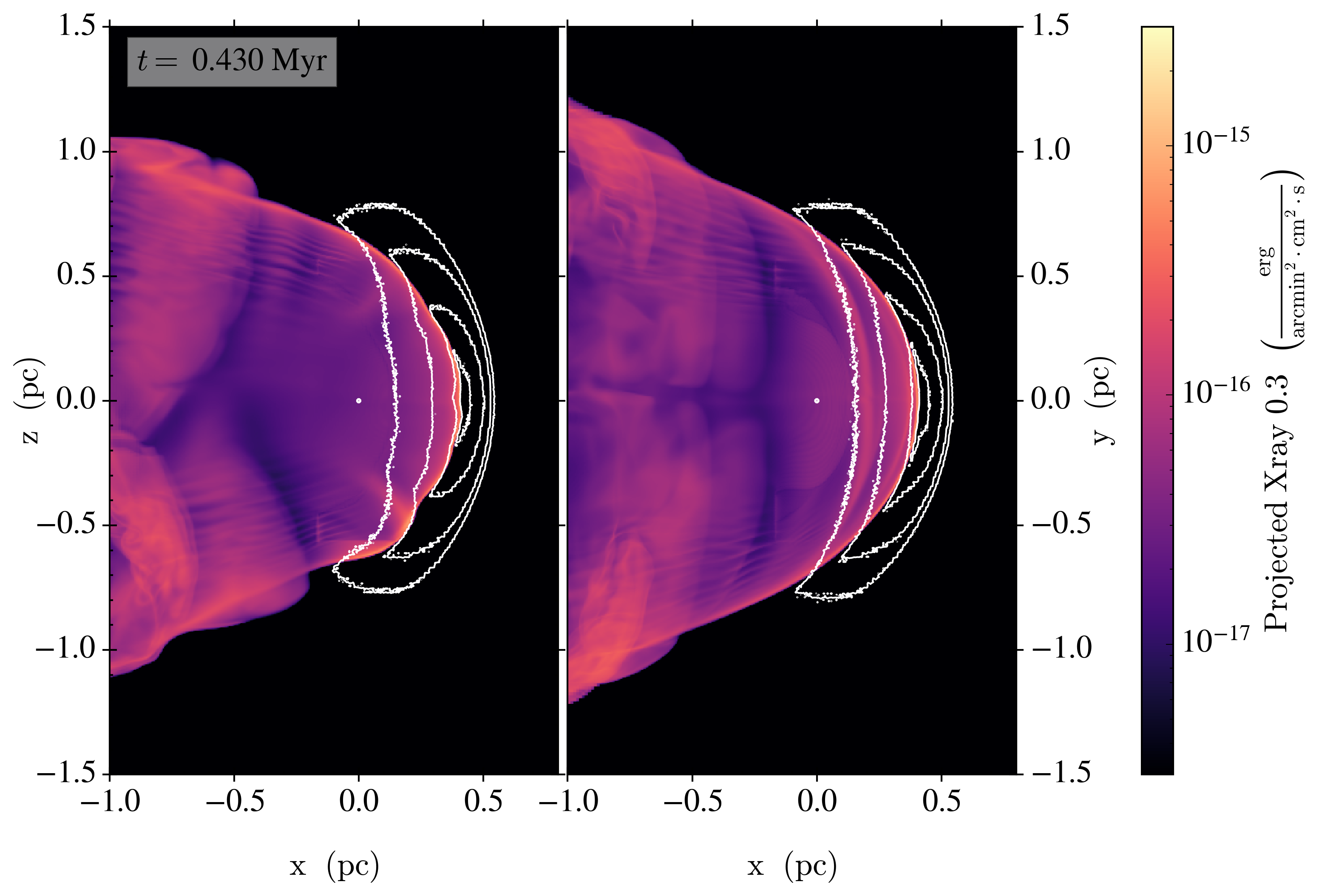}
    \caption{X-ray and $24\,\mu$m IR emission from simulation \texttt{3D-3-by7} at two different simulation times as indicated in the plot labels.
            Contours of IR brightness are at a levels of 50,100,200 and 300\,MJy\,ster$^{-1}$, with the peak at the apex of the bow shock.}
    \label{fig:3d-mhd-xrayir}
\end{figure*}

\subsection{3D MHD simulations: Numerical resolution and magnetic field orientation}
Fig.~\ref{fig:3d-mhd-bx-res-sametime} shows slices of gas density through the planes $y=0$ and $z=0$ for 3D simulations with the ISM magnetic field aligned with the space velocity of the star, that is, the same configuration as the 2D MHD simulations, and using the HLLD/HLL hybrid solver.
This shows very similar results to those obtained with the equivalent resolution 2D simulation, namely that low resolution ($128^3$, top row) shows very weak instability at the CD and a generally smooth flow, medium resolution ($256^3$, middle row) shows some large-scale fluctuations on the CD driven by poorly resolved KH instability, and high resolution ($384^3$, bottom row) shows well-developed vortices and significant mixing of ISM material into the shocked wind region.
Higher resolutions were not feasible with the computational resources we had available for this project.

Fig.~\ref{fig:3d-mhd-by-res-sametime} shows the same but for the 2nd magnetic field configuration, $\bm{B_0}=[1,4,0]\,\mu\mathrm{G}$.
Here we expect the CD may be more stable because the ISM field lines are draped across the CD, based on MHD simulations of the Heliosphere \citep{PogZanOgi04} and of bow shocks produced by runaway red-supergiant stars \citep{MeyMigPet21}.
2D simulations of the KH instability in the presence of a magnetic field also show strong inhibition of the growth rate for fields parallel to the discontinuity \citep{FraJonRyu96, RyuJonFra00}.
This is true up to a point, in that the lower resolution cases ($128^3$ and $256^3$) show only small perturbations on the CD, which is dominated by the strong shear flow and maintains a relatively stable configuration.

At the highest resolution that we could achieve ($384^3$, bottom panel) this is no longer the case.
In the $x$-$y$ plane the CD is quite stable near the apex of the bow shock (bottom-right panel), because here the ISM magnetic field is entirely in the image plane and parallel to the shear flow across the CD.
In the $x$-$z$ plane, perturbations grow as they move outwards because here the magnetic field is almost perpendicular to this plane and so the instability behaves rather like in the case of zero field \citep{FraJonRyu96}.
The KH instability can grow to the non-linear phase \citep[e.g.][]{JunWalHei10}, resulting in significant entrainment of ISM material into the wake behind the bow shock.
This mixing is driven by vortices developing in flows perpendicular to the ISM magnetic field (i.e.\ in the $\pm\hat{z}$ direction), but the mixed material fills the turbulent wake behind the star.
These results arise because the more accurate hybrid HLLD/HLL solver, together with increased numerical resolution, allows shorter wavelength KH modes (with shorter growth timescales) to be captured.
The solution has moved from a regime dominated by numerical diffusion (lower-resolution cases) to a regime dominated by hydrodynamic mixing processes in the highest-resolution case.

\subsection{3D MHD simulations: Time dependence of X-ray emission}

Fig.~\ref{fig:xray-lum-3d-mhd} shows the X-ray luminosity in the 0.3-10\,keV band for the suite of 3D MHD simulations for ISM magnetic fields in the parallel (a) and near-perpendicular (b) configurations.
Very similar features are seen, when compared with 2D results using the HLLD/HLL solver: the initial expansion phase that lasts for about 0.1\,Myr, followed by a quasi-periodic variation on approximately the same timescale.
Whereas in 2D the amplitude of the slow variation increased with resolution, saturating at an amplitude of a factor of 10-20, in 3D for the B-field-aligned case (a), the amplitude is not obviously increasing with resolution, but stays at about a factor of 5-10 for the three resolutions that we could simulate (although at the highest resolution the last few snapshots show a higher peak in X-ray luminosity).

For the perpendicular-field case (b), the fluctuation is strongly increasing with resolution, from a factor of 2 at the lowest resolution to a factor of 10 at a resolution 4 times larger.
The computational cost of simulations \texttt{3D-3-bx7} and \texttt{3D-3-by7} means that we could not reach a runtime of 0.5\,Myr, but the trend of increasing variability with increasing resolution is clear for the perpendicular-field case.

Fig.~\ref{fig:3d-mhd-by-comp1} shows a snapshot of simulation \texttt{3D-3-by7} at $t=0.39$\,Myr, plotting gas density slices, X-ray emission maps and magnetic field slices from top to bottom.
At this time the X-ray luminosity of the simulation is $L_\mathrm{X}\approx3\times10^{29}$\,erg\,s$^{-1}$, approximately the mean value.
The density slices show that significant entrainment of ISM gas into the wake behind the star has occured, and relatively large deformations of the contact discontinuity of the bow shock are apparent in the $x$-$z$ plane.
The outline of these deformations are apparent in the projected X-ray emission maps, and the X-ray emission fluctuates strongly in the downstream region (note that the emission map uses a logarithmic scale).
The magnetic field is strongly compressed in the bow shock because of the near-perpendicular configuration of the ISM field, reaching values above 20\,$\mu$G.
In the shocked stellar wind region the magnetic field is sub-$\mu$G, with thin regions of very weak field from the poles of the star (where the toroidal component is near zero) and from the equatorial current sheet.
In regions of strong wind-ISM mixing, the magnetic field tends to be much stronger.

The same plots are shown in Fig.~\ref{fig:3d-mhd-by-comp2} at a later time of $t=0.43$\,Myr, near the time of maximum X-ray luminosity.
Here, the largely deformed CD that was noted at $t=0.39$\,Myr has resulted in turbulent mixing of wind and ISM gas in the downstream wake at $x\lesssim -1.5$\,pc.
The CD near the apex of the bow shock now has only weak deformations.
The X-ray emission maps show again a strongly fluctuating emission as a function of position, but are almost everywhere a factor of a few brighter than in Fig.~\ref{fig:3d-mhd-by-comp1}.
The brightest X-ray emission is in the downstream wake behind the star, as was found by \citet{GreMacHaw19} for 2D HD simulations.
The magnetic field in the turbulent mixing region is chaotic and comparatively stronger than at $t=0.39$\,Myr, in most regions with a strength of a few $\mu$G.

Fig.~\ref{fig:phase-3d-mhd-by} shows the volume fraction of gas in the $\rho-T$ plane for simulation \texttt{3D-3-by7}, plotted at times $t=0.35$\,Myr, 0.39\,Myr and 0.43\,Myr.
The gas at the lower-left part of the plane is the unshocked stellar wind, for which the thermal pressure (and hence $T$) is not accurately modelled on account of the strongly kinetic-energy-dominated flow (only total energy is conserved by the MHD solver).
The bulk of the gas volume is at the bottom-right corner of the plane, photoionised ISM at the equilibrium temperature of $T\sim10^4$\,K.
The next highest fraction of the gas volume is at the upper-left corner of the plane: the hot and tenuous shocked wind.
The gas connecting the shocked wind to the photoionised ISM is the mixing region, with $4.5\lesssim \log_{10} T / \mathrm{K} \lesssim 6.5$, which is isobaric to a very good approximation (the solid black line shows gas at constant pressure).
The mean pressure of this gas in the mixing layer increases from top to bottom, correlating well with the increase in X-ray luminosity.

The differential emission measure, or DEM, is often used to show the X-ray emission properties of simulations \citep[e.g.][]{ToaArt18, GreMacHaw19}.
The DEM from simulations \texttt{3D-1-by7}, \texttt{3D-2-by7} and \texttt{3D-3-by7} are plotted at selected times in Fig.~\ref{fig:dem-3d-mhd-by}.
At the lowest resolution (top panel) there is very little time variation, and this is reflected in the relatively flat X-ray luminosity evolution (Fig.~\ref{fig:xray-lum-3d-mhd}).
The degree of variation increases significantly in the medium resolution (middle panel) and further for the highest resolution (bottom panel) simulation.
Here we see that at the X-ray emitting temperatures ($T\sim10^6-10^{7.5}$\,K for the 0.3-2\,keV band) the DEM varies by more than an order of magnitude from minimum to maximum, with the maximum at $t=0.43$\,Myr and minimum values corresponding to the minima of the X-ray light curve (Fig.~\ref{fig:xray-lum-3d-mhd}).
The slope of the DEM gets shallower at times of stronger X-ray emission, indicating a larger (but still small) fraction of the radiative emission will emerge at X-ray energies.

\subsection{IR emission from dust and relation to X-ray emission}
 
As described in section \ref{sec:methods:postprocess}, we generated maps of thermal-infrared (IR) dust emission from our simulations using the \textsc{torus} radiation-hydrodynamics code in post-processing mode.
Fig.~\ref{fig:3d-mhd-ir} shows a sequence of snapshots from simulation \texttt{3D-3-by7} at wavelengths 24 and 70\,$\mu$m, corresponding to the central wavelengths of the \textit{Spitzer Space Telescope MIPS} and \textit{Herschel PACS} instruments.
As with the X-ray emission maps and density plots, the bow shock appears very smooth in the $x$-$y$ plane because there the ISM magnetic field is draped along the bow shock and contact discontinuity, strongly inhibiting the development of dynamical instabilities.
In the $x$-$z$ plane, however, the development of large-scale deformations of the bow shock is apparent at both wavelengths.
These results imply that IR emission may be used to study the stability of the wind-ISM interface in bow shocks, but that a smooth bow shock in observations may in fact be quite distorted if viewed from a different perspective, depending on the relative orientation of the ISM magnetic field.

At 24\,$\mu$m the dust emission is very bright near the apex of the bow shock and decreases rapidly with distance from the star \citep{AcrSteHar16, MacHawGva16} because the dust temperature is low enough that the Wien law ensures that a small decrease in temperature produces an exponential decrease in 24\,$\mu$m emissivity.
The dependence on distance is much weaker at 70\,$\mu$m because closer to the peak emission wavelength the emissivity is less sensitive to temperature.
\citet{AcrSteHar16} considered a more massive star with a stronger wind and moving into a much denser ISM than in our simulations; consequently their bow shock is closer to the more luminous star and so their dust emission peaks at shorter wavelength than in our case.

Fig.~\ref{fig:3d-mhd-ir} shows that the outer (forward) shock is well traced by a sharp decrease in $24\,\mu$m and $70\,\mu$m IR brightness with increasing distance from the star.
The brightest region of mid-IR emission is between the CD and the forward shock, as expected because the dust density is largest in this region and the radiative heating is strongest.
Although somewhat smeared out by projection effects, the CD is still well traced by a strong gradient of increasing mid-IR emission with increasing distance from the star.
Confirming this finding, Fig.~\ref{fig:3d-mhd-xrayir} shows the same two snapshots as Figs.~\ref{fig:3d-mhd-by-comp1} and \ref{fig:3d-mhd-by-comp2}, with projected X-ray brightness over-plotted with $24\,\mu$m IR emission contours at 50, 100, 200 and 300\,MJy\,ster$^{-1}$.
The X-ray emission obviously traces the contact discontinuity very well because of the sharp temperature gradient.
There is significant overlap between fainter IR emission and X-ray emission, which is a projection effect of the curved surface of the bow shock.
Nevertheless the brightest IR emission is anti-correlated with the X-ray emission, and located upstream from the edge of the X-ray-emitting volume.

\section{Discussion}
\label{sec:discussion}

As described above in Section~\ref{sec:introduction}, previous simulations that calculated X-ray emission from bow shocks showed some seemingly contradictory results when compared with the \textit{Chandra} detection of diffuse X-ray emission around $\zeta$ Oph \citep{ToaOskGon16, GreMacKav22} and the non-detection of the Bubble Nebula by \textit{XMM-Newton} \citep{GreMacHaw19, ToaGueTod20}.
It was speculated that numerical effects such as limitations imposed by 2D axisymmetric calculations \citep{GreMacHaw19} or lack of resolution to capture KH instability in 3D \citep{GreMacKav22} could result in predicted X-ray emission being too high or too low, respectively.
Furthermore, the differences between HD and MHD simulations and the simplifications of modelling a star moving through a uniform medium could also play a role, especially for the Bubble Nebula where the star is clearly moving into a dense and clumpy environment \citep{ToaGueTod20}.
We have addressed some of these uncertainties by simulating the same bow shock many times with different dimensionality (2D and 3D), resolution (spanning a factor of 8 in cell diameter), equations solved (HD and MHD) and type of solver (HLL and HLLD).

We found that dimensionality and the presence of magnetic fields appear to have little effect on the results, at least for the values of $\rho_0$, $\bm{B}_0$ and $v_\star$ considered here.
This is primarily because the apex of the bow shock and the layer of shocked ISM are well-resolved and stable, and not subject to the thin-shell instability \citep{Vis94, BloKoe98}, which is difficult to model properly in 2D on account of the coordinate singularity at the symmetry axis.
For larger $v_\star$, the thin-shell instability will play a stronger role and 2D simulations may no longer be feasible \citep{GreMacHaw19}.
Where 3D does make a noticeable difference is for synthetic emission maps, which are more realistic in 3D because the symmetry assumed in 2D calculations results in artificial rings of emission \citep{MeyMacLan14, AcrSteHar16, GreMacHaw19}.
This enables predictions for X-ray emission morphology that may be directly tested by future observations.

Numerical resolution and the choice of flux solver play a large role in the results obtained.
In particular, the HLL solver cannot resolve CDs and reduces the strong density and temperature gradients by numerical diffusion, thereby limiting the development of KH instability.
This was noted in the context of colliding-wind flows in binary systems by \citet{LamFroDub11}.
At resolutions of $128\times64$ or $256\times128$ and using the HLL solver, the development of KH instabilities is not captured well and numerical diffusion produces a relatively smooth layer of intermediate density and temperature.
The HLLD solver does much better at resolving the CD and so KH waves are more apparent.
For X-ray emission from the mixing layer, the HLL solver showed a systematically decreasing emission with increasing resolution, as well as stronger variability.
The hybrid solvers also showed stronger variability with increasing resolution, but there was no apparent overall trend of luminosity changing with resolution.
These results agree with the numerical experiments of \citet{TanOhGro21}, who find that the emissivity of shear-mixing layers should decrease with increasing resolution up to the point at which the turbulent mixing (i.e.\ entrainment) takes over from diffusive mixing as the dominant process in the mixing layer.

There are some other notable agreements with previous work on turbulent mixing layers.
\citet{FieDruOst20} also find that the turbulent mixing layer in their simulations is isobaric when sufficiently resolved (see Fig.~\ref{fig:phase-3d-mhd-by}), interpreting this as evidence for mixing driven by turbulence rather than by runaway cooling (i.e.\ cooling flows).
\citet{TanOhGro21} identify the eddy turnover timescale $L/u^\prime$ as a key parameter (where $L$ was the size of the largest eddies and $u^\prime$ the characteristic turbulent velocity), particularly the ratio of this timescale to the cooling time in the mixing layer.
The equivalent timescale in our simulations is the advection timescale along the mixing layer from the star to the downstream boundary, $\tau_\mathrm{ad} = (x_\star - x_\mathrm{min}) / v_\star =0.116$\,Myr (although there is some ambiguity in defining the turbulent velocity for a multi-phase medium).
We see significant variation in emissivity of the mixing layer on this timescale, driven by the growth of large eddies from the stagnation point downstream to when they leave the simulation domain.

\citet{BaaSchKle21, BaaSchKle22} made 2D and 3D MHD investigations of bow shocks from hot stars, although they only investigated one field orientation, being mainly interested in the effects of ISM inhomogeneities and of varying $v_\star$ on the properties of the bow shock and astrosphere.
The 3D simulations of \citet{BaaSchKle22} show some oscillation in the position of the Mach Disk, but do not show any dynamical instabilities arising at the CD, and so the mixing layer between wind and ISM is laminar and presumably regulated by numerical diffusion.
Their numerical setup also uses the diffusive HLL solver and a second-order-accurate integration scheme, and so it is not surprising that a laminar CD is seen, similar to the results we have obtained with the HLL solver.
To correctly capture the dynamics of the shear layer near the CD, either higher spatial resolution or a more accurate Riemann solver should be used.

Similarly, MHD simulations of bow shocks with \textsc{pluto} \citep{MeyMigKui17, MeyMigPet21} used second-order methods with the HLL solver, for which we have shown that the CD and its stability are poorly modelled.
The \textsc{pluto} MHD code has a number of high-order schemes \citep{MigBerRos24} that may be sufficiently robust to apply to the bow-shock problem, and it may be more profitable to apply higher-order integration methods to obtain better resolution at the CD with less computational effort.
For example the \textsc{gamera} MHD code \citep{ZhaSorLyo19} has up to 8th-order spatial accuracy and has been used to model planetary magnetospheres at very high resolution, including resolving waves forming at the CD \citep{SorMerPan20}, albeit with a weaker CD than is present in bow shocks from massive stars.

One of our main results is that both local and global X-ray emission from the shocked wind is highly variable, by at least an order or magnitude, depending on the instantaneous degree of mixing that is occurring in the wake behind the star.
Mixing is driven by the development of waves on the CD surface, namely, the KH instability arising from the large velocity shear across the CD.
The global X-ray variability that we measure is to some extent an artefact of the limited spatial domain and X-ray-bright regions being advected downstream out of the domain, and indeed the variability timescale matches very closely the advection timescale from the star to the downstream boundary.
Nevertheless, pointed X-ray observations also have a limited field of view and cannot probe the full extent of the turbulent wake downstream from the star \citep{ToaOskGon16, ToaGueTod20}, and so it is reasonable to consider the total emission within a limited volume.
The local variability at any given point downstream from the star also undergoes similarly large variations in surface brightness (compare Figs.~\ref{fig:3d-mhd-by-comp1} and~\ref{fig:3d-mhd-by-comp2}), and so the X-ray surface brightness measured by an observation should also vary depending on the instantaneous properties of the mixing layer.

Large X-ray variability is obtained for the case of a uniform ISM, for $v_\star$ low enough that the bow shock does not experience the thin-shell instability, arguably the most stable possible conditions.
One may expect that, for a structured ISM with overdense clumps and turbulent motions, larger variations in X-ray emission will occur due to the variable upstream ram-pressure.
Similarly, for larger $v_\star$ the compression ratio of the forward shock will increase and the shell will be increasingly subject to the thin-shell instability, again imposing large-scale perturbations on the CD \citep{DgaBurNor96, BloKoe98} that one may expect to have a large effect on the X-ray luminosity of the hot bubble.

\section{Conclusions}
\label{sec:conclusions}

We have undertaken an in-depth numerical study of bow shocks around runaway massive stars using 2D and 3D MHD simulations, primarily to investigate thermal X-ray emission from the shocked stellar wind.
Our findings are as follows:
\begin{itemize}
\item For low-resolution simulations in 2D and 3D the CD remains laminar, mixing is dominated by numerical diffusivity, and X-ray emission from the interface is resolution dependent.
\item With sufficient spatial resolution and when using a solver that accurately captures the CD, the KH instability develops at the CD due to the strong shear between the wind and ISM gas, in both HD and MHD simulations.
\item Large-scale eddies developing from the KH instability entrain dense ISM gas into the hot stellar wind, producing fluctuations in X-ray luminosity from the hot gas on a timescale $\tau\approx L_\mathrm{box} / v_\star$, with a ratio between maximum and minimum luminosity of up to $20\times$.
\item X-ray surface brightness also has strong spatial variations, peaking downstream where the mixing layer has produced gas at intermediate densities and temperatures that emit strongly in soft X-rays.
\item Although $\lesssim 10^{-4}$ of the input mechanical energy of the bow shock is radiated at X-ray energies, the X-ray luminosity is nevertheless a good tracer of the degree of mixing at the wind-ISM interface.
\item the CD appears smooth from viewing angles where the (here uniform) ISM magnetic field is mostly in the image plane, because the draping of field lines along the bow shock inhibits the development of KH instability in this direction.
\item In the perpendicular direction the flow behaves like a hydrodynamic flow and distortion of the CD is visible in synthetic IR dust emission maps and X-ray surface-brightness maps.
\item There is a strong anti-correlation between X-ray and IR surface brightness, with some overlap due to projection effects.  The CD is located where the IR intensity gradient is strongest.
\item Despite the different geometry and simulation setup, there appears to be a strong correspondence between our results and recent work exploring dissipation in turbulent shear layers.  This should be explored further in more detail in future work.
\end{itemize}

\begin{acknowledgements}
We acknowledge the referee for a careful reading of the manuscript and useful suggestions to improve it.
This work was supported by an Irish Research Council (IRC) Starting Laureate Award.
JM acknowledges support from a Royal Society-Science Foundation Ireland University Research Fellowship.
AM acknowledges support from a Royal Society Research Fellows Enhancement Award 2021.
TJH acknowledges funding from a Royal Society Dorothy Hodgkin Fellowship and UKRI guaranteed funding for a Horizon Europe ERC consolidator grant (EP/Y024710/1).
RB acknowledges funding from the Irish Research Council under the Government of Ireland Postdoctoral Fellowship program.
This publication results from research conducted with the financial support of Taighde \'Eireann - Research Ireland under Grant numbers 20/RS-URF-R/3712, 22/RS-EA/3810, IRCLA\textbackslash 2017\textbackslash 83.
SW acknowledges the support of the Deutsche Forschungsgemeinschaft (DFG) via the Collaborative Research Center SFB 1601 `Habitats of massive stars across cosmic time' (subprojects A5 and B5).
This research made use of the following software: VisIt \citep{VisIt}, Astropy \citep{astropy:2018},  Numpy \citep{HarMilVan20}, matplotlib \citep{Hun07}, yt \citep{TurSmiOis11}, \textsc{pion} \citep{MacGreMou21}, \textsc{pypion} \citep{GreMac21}, \textsc{heasoft} \citep{HEAsoft14}, \textsc{Xspec} \citep{Arn96}.
\end{acknowledgements}


\bibliographystyle{aa}
\bibliography{refs}


\end{document}